\documentclass[twocolumn]{aastex701}
\newcommand{\kms}{km~s$^{-1}$}
\newcommand{\cmcube}{cm$^{-3}$}
\newcommand{\nbar}{\bar{n}_{\rm H_2}}
\newcommand{\tdyn}{t_{\rm dyn}}
\newcommand{\ntdyn}{\bar{n}_{\rm H_2} t_{\rm dyn}}
\newcommand{\sigv}{\sigma_v}

\shorttitle{Nobeyama 45~m CMZ Survey I} 
\shortauthors{Takekawa et al.}

\usepackage{needspace}

\begin{document}

\title{The Nobeyama 45~m Survey of Shocked Molecular Gas in the Central Molecular Zone. I. Survey Data, Cloud Catalog, and SiO Line-Ratio Trends}

\author[0000-0001-8147-6817]{Shunya Takekawa}
\email{shunya@kanagawa-u.ac.jp}
\affiliation{Faculty of Engineering, Kanagawa University, 3-27-1 Rokkakubashi, Kanagawa-ku, Yokohama, Kanagawa 221-8686, Japan}
\author[0000-0002-1663-9103]{Shiho Tsujimoto}
\email{shiho.tsujimoto@keio.jp}
\affiliation{School of Fundamental Science and Technology, Graduate School of Science and Technology, Keio University, 3-14-1 Hiyoshi, Kohoku-ku, Yokohama, Kanagawa 223-8522, Japan}
\author[0000-0002-5566-0634]{Tomoharu Oka}
\email{tomo@phys.keio.ac.jp}
\affiliation{Department of Physics, Institute of Science and Technology, Keio University, 3-14-1 Hiyoshi, Kohoku-ku, Yokohama, Kanagawa 223-8522, Japan}
\affiliation{School of Fundamental Science and Technology, Graduate School of Science and Technology, Keio University, 3-14-1 Hiyoshi, Kohoku-ku, Yokohama, Kanagawa 223-8522, Japan}
\author[0000-0003-2735-3239]{Rei Enokiya}
\email{rei.enokiya@nao.ac.jp}
\affiliation{Chile Observatory, National Astronomical Observatory of Japan, National Institutes of Natural Sciences, 2-21-1 Osawa, Mitaka, Tokyo 181-8588, Japan}

\begin{abstract}
We present large-scale molecular line maps of the Central Molecular Zone (CMZ) in our Galaxy obtained with the Nobeyama Radio Observatory 45~m telescope.
The observations cover a $3\fdg5 \times 0\fdg5$ region with 20\arcsec\ resolution in eight molecular lines including SiO $J$=2--1, CS $J$=2--1, H$^{13}$CN $J$=1--0, and HCN $J$=1--0.
We release the calibrated data cubes and a catalog of SiO-emitting clouds identified by the SCIMES
algorithm. 
For each cloud, we provide cloud-integrated intensities of the observed lines and derive an H$^{13}$CN-based molecular gas mass and a cloud-averaged H$_2$ number density.
We compare SiO intensity ratios with the density, dynamical time, and their product $\ntdyn{}$.
We find that SiO intensity ratios relative to six other molecular lines all decrease systematically with increasing $\ntdyn{}$.
The tightest correlations are found for SiO/H$^{13}$CN ($r_s = -0.82$) and SiO/CS ($r_s = -0.70$).
The released data cubes and cloud catalog enable systematic studies of SiO enhancement, shock chemistry, and gas dynamical evolution across the CMZ.
\end{abstract}


%% ============================================================
\section{Introduction} \label{sec:intro}
The Central Molecular Zone (CMZ) is the most extreme molecular gas environment in our Galaxy, extending within a radius of $\sim$250~pc from the Galactic nucleus Sgr~A$^*$.
Molecular gas in the CMZ is characterized by high density ($n_{\rm H_2} \gtrsim 10^4$ cm$^{-3}$), high temperatures ($T_{\rm k} \gtrsim 50$ K), and large velocity dispersions ($\sigma_v \gtrsim 10$ \kms{}) \citep[e.g.,][]{morris96, oka98, oka12, ginsburg16, tanaka18}.
These physical properties are accompanied by numerous shock signatures, particularly the extensive emission from shock-tracing molecules such as SiO and HNCO, as well as collisionally excited CH$_3$OH and OH masers \citep[e.g.,][]{martinpintado97, huttemeister98, wardle02, requena-torres06, riquelme10sio, yusef-zadeh13}.
The origin of these widespread shocks remains debated. Possible mechanisms include large-scale gas flows driven by the Galactic bar \citep[e.g.,][]{kruijssen15, sormani18}, cloud--cloud collisions \citep[e.g.,][]{hasegawa94, tsuboi15, enokiya22}, and past nuclear activity of Sgr~A$^*$ \citep{takekawa24}.
While these global mechanisms may drive widespread shocks, compact disturbed features such as high-velocity compact clouds (HVCCs) are associated with strong local shocks \citep{tanaka14, tanaka15, oka16, takekawa17, takekawa19a, takekawa19b, takekawa20}.

SiO is one of the most widely used tracers of shocked molecular gas.
In quiescent molecular clouds, silicon is mostly depleted onto dust grains, whereas shocks can liberate silicon-bearing material and enhance gas-phase SiO by several orders of magnitude \citep[e.g.,][]{ziurys89, schilke97}.
This enhancement is temporary because gas-phase SiO depletes back onto dust grains on timescales of $\sim 10^4$--$10^5$~yr at typical CMZ densities ($n_{\rm H_2} \sim 10^4$--$10^5$ cm$^{-3}$; \citealt{martinpintado92, bergin98}).
Consequently, in high-density environments like the CMZ, strong SiO emission effectively highlights recent shock events.

Previous single-dish studies have mapped SiO emission across the CMZ or sampled selected molecular peaks, establishing the large-scale distribution of shocked gas \citep{martinpintado97, huttemeister98, riquelme10sio, jones12, tsuboi15}.
These data differ mainly in mapped area, angular resolution, and line coverage.
\citet{martinpintado97} mapped SiO $J$=1--0 over a
$1\arcdeg \times 0\fdg2$ region at $2'$ resolution with the
Yebes 14~m telescope and followed up selected Sgr~A and
Sgr~B fields in SiO $J$=2--1 at $26''$ resolution with
the IRAM 30~m telescope.
The NANTEN survey covered a much wider Galactic center region at $\sim 3\farcm6$ resolution \citep{riquelme10sio}.
The Mopra 3~mm survey mapped $2\fdg5 \times 0\fdg5$ in many molecular lines at $\sim 40\arcsec$ resolution \citep{jones12}.
Previous Nobeyama Radio Observatory (NRO) 45~m SiO and H$^{13}$CO$^+$ observations covered the central $1\fdg5 \times 0\fdg5$ region with an effective $42\arcsec$ beam, including $26\arcsec$ data toward Sgr~A \citep{tsuboi15}.
More recently, the Atacama Large Millimeter/submillimeter Array (ALMA) CMZ Exploration Survey (ACES) has resolved intricate structures at arcsecond resolution within a Galactocentric radius of $\sim 100$~pc in many dense-gas and shock-sensitive tracers \citep{longmore_aces, walker_aces, lu_aces, hsieh_aces}.
While these ALMA data reveal the internal structure of molecular clouds,
single-dish mapping over the full CMZ provides the cloud-scale distribution
of dense and shocked molecular gas.

To systematically investigate the shocked molecular gas across the CMZ, we conducted large-scale molecular line imaging observations with the Nobeyama 45~m telescope.
The resulting SiO map covers $3\fdg5 \times 0\fdg5$ with a 20\arcsec\ beam.
It covers a wider longitude range and has a finer spatial resolution than the Mopra 3~mm survey and the previous NRO 45~m SiO maps.
The survey also includes seven reference molecular lines, allowing SiO emission to be compared with dense-gas and shock-sensitive tracers over the same field.
Parts of the survey data have been used in studies of individual sources \citep{tsujimoto21, kaneko23}, and \citet{takekawa24} presented an initial survey-wide analysis of the SiO line ratios. 

Here, we present the full calibrated data products, integrated maps, velocity-channel maps, the SiO-emitting cloud catalog, and cloud-integrated line intensities.
Using the derived catalog, we investigate the dependence of SiO intensity ratios on the cloud-averaged density and dynamical time.
More detailed chemical interpretation of these global trends and in-depth analyses of specific structures such as HVCCs will be addressed in subsequent papers.

Section~\ref{sec:obs} describes the observations, data reduction, and released data products.
Section~\ref{sec:results} presents the survey-scale integrated-intensity and velocity-channel maps.
Section~\ref{sec:cloudcatalog} presents the SiO-emitting cloud catalog, derived cloud properties, and line-ratio trends.
Section~\ref{sec:summary} summarizes the main conclusions.
Throughout this paper, we assume a Galactic center distance of 8.28~kpc \citep{gravity22}.

%% ============================================================
\section{Observations and Data Reduction} \label{sec:obs}

\subsection{Observations}

We mapped eight molecular lines over nearly the full CMZ with the Nobeyama Radio Observatory (NRO) 45~m telescope.
The survey covers a $3\fdg5 \times 0\fdg5$ area, spanning Galactic longitude $-1\fdg5 \leq l \leq +2\fdg0$ and latitude $-0\fdg25 \leq b \leq +0\fdg25$.
The main survey region, covering $-1\fdg5 \leq l \leq +1\fdg5$, was observed under the NRO 45~m Large Program LP187001 (PI: S.~Takekawa) in 2019 January--May, 2020 January--April, and 2021 January--April.
The eastern extension, $+1\fdg5 \leq l \leq +2\fdg0$, was added in 2023 January--February through program G22022 (PI: S.~Takekawa).
The total telescope time was 310~hr.

Observations were made in On-The-Fly (OTF) mode \citep{sawada08} using the two-sideband receiver FOREST \citep{minamidani16} and the SAM45 spectrometer \citep{kamazaki12}.
The eight molecular transitions listed in Table~\ref{tab:obs} were covered by two frequency setups.
Setup~1 includes SiO $J$=2--1, CS $J$=2--1, H$^{13}$CN $J$=1--0, H$^{13}$CO$^+$ $J$=1--0, and CH$_3$OH $J_{K_a,K_c}$=2$_{1,1}$--1$_{1,0}$~A$^-$, while Setup~2 covers HCN $J$=1--0, HCO$^+$ $J$=1--0, and SO $N_J$=2$_3$--1$_2$.
The corresponding local oscillator (LO) frequencies were 92.3~GHz for Setup~1 and 94.0~GHz for Setup~2. Both setups were configured with a 1~GHz spectrometer bandwidth and a channel spacing of 244.14~kHz.
For Setup~1, OTF scans were performed along both Galactic longitude (x-scan) and latitude (y-scan), while Setup~2 was observed only in the Galactic-longitude direction (x-scan).
The mapping was performed by dividing the survey region into rectangular tiles.
In the main region ($|l| \leq 1\fdg5$), x-scans covered $36\arcmin \times 10\arcmin$ tiles and y-scans covered $30\arcmin \times 10\arcmin$ tiles. This required 15 x-scan tiles and 18 y-scan tiles to fill the main region.
For the eastern extension ($l > 1\fdg5$), both x- and y-scans covered $30\arcmin \times 10\arcmin$ tiles, requiring 3 tiles each.
The x-scans consisted of 121 rows with a $5\arcsec$ spacing. A scan duration of 16~s per row with a data-dump time of 0.04~s corresponds to a scan speed of $135\arcsec$~s$^{-1}$, and a sampling interval of $5\farcs4$ along the scan direction.
The y-scans similarly used a $5\arcsec$ row spacing and a scan duration of 14~s per row with the same dump time.
This corresponds to a scan speed of $128\farcs6$~s$^{-1}$ and a $5\farcs1$ sampling interval.
Pointing accuracy was checked roughly every 1.5~hr using the SiO maser VX~Sgr at 43~GHz with the H40 receiver, keeping the pointing uncertainty within $5\arcsec$.

\begin{deluxetable}{lcccc}
\tablenum{1}
\tablecaption{Observed Molecular Lines \label{tab:obs}}
\tablewidth{0pt}
\tablehead{
\colhead{Molecule} & \colhead{Transition} & \colhead{Frequency} & \colhead{$\eta_{\rm MB}$\tablenotemark{a}} & \colhead{$\Delta T_{\rm MB}$\tablenotemark{b}} \\
\colhead{} & \colhead{} & \colhead{(GHz)} & \colhead{} & \colhead{(K)}
}
\startdata
\multicolumn{5}{c}{Setup 1 (LO frequency of 92.3 GHz)} \\
\hline
H$^{13}$CN & $J$=1--0 & 86.3399 & 0.50 & 0.14 \\
H$^{13}$CO$^+$ & $J$=1--0 & 86.7543 & 0.50 & 0.14 \\
SiO & $J$=2--1 & 86.8470 & 0.50 & 0.13 \\
CH$_3$OH & $J_{K_a,K_c}$=2$_{1,1}$--1$_{1,0}$~A$^-$ & 97.5828 & 0.45 & 0.15 \\
CS & $J$=2--1 & 97.9810 & 0.45 & 0.15 \\
\hline
\multicolumn{5}{c}{Setup 2 (LO frequency of 94.0 GHz)} \\
\hline
HCN & $J$=1--0 & 88.6316 & 0.50 & 0.20 \\
HCO$^+$ & $J$=1--0 & 89.1885 & 0.50 & 0.21 \\
SO & $N_J$=2$_3$--1$_2$ & 99.2999 & 0.45 & 0.20 \\
\enddata
\tablenotetext{a}{Main-beam efficiency.}
\tablenotetext{b}{The rms noise level measured in the data cube with a $7\farcs5 \times 7\farcs5 \times 2$~\kms{} grid.}
\end{deluxetable}

\subsection{Calibration}

The antenna temperature scale $T_{\rm a}^*$ was calibrated using the standard chopper-wheel method.
In the 2019 observing runs, we used three reference off positions at $(l, b) = (+1\fdg0, -0\fdg5)$, $(0\fdg0, -0\fdg5)$, and $(-1\fdg0, -0\fdg5)$.
One of these, $(0\fdg0, -0\fdg5)$, was later found to contain CS emission. This contamination produced an artificial absorption-like feature near $V_{\rm LSR} \simeq 19$~\kms{} in the affected CS spectra.
For observations after 2020, we shifted the off positions to $b = -0\fdg75$.
For the CS line, we performed $\sim$10 minute integrations toward the new off positions and confirmed that the off-position spectra showed no significant emission in any of the four beams with typical $T_\mathrm{A}^*$ rms noise levels of $\sim 0.02$ K.
The contaminated 2019 CS spectra were corrected by modeling the off-position emission. Details of this procedure are given in Appendix~\ref{sec:offpos}.

To monitor the day-to-day stability of the intensity scale, we mapped a $3\arcmin \times 3\arcmin$ field centered on $(l, b) = (0\arcdeg\,4\arcmin\,0\arcsec, -0\arcdeg\,2\arcmin\,0\arcsec)$ in the Sgr~B2 region once per observing day.
From these observations, we derived daily scaling factors.
The CS line was used for Setup~1, and the HCN line was used for Setup~2.
We selected 2019 January 19 and 2020 March 26 as the reference days for Setup~1 and Setup~2, respectively, because they provided the best observing conditions with good weather and low system temperatures.
For each setup, the integrated intensity of the corresponding line in the average spectrum of the central part of this field on the reference day was set to unity.
The scaling factors were derived for each observing day and applied to the data.
The derived factors indicated that the intensity scale was stable within about 10\% on most days.

\subsection{Data Reduction and Data Products}

We reduced the data with the Nobeyama OTF Software Tools for Analysis and Reduction (NOSTAR) package provided by NRO \citep{sawada08}.
After baseline subtraction, the spectra were gridded onto data cubes sampled at $7\farcs5 \times 7\farcs5 \times 1$~\kms{} using a Bessel--Gaussian convolution kernel.
For Setup~1, separate x-scan and y-scan maps were first produced and then combined using the basket-weaving method \citep{emerson88} to reduce scanning effects.
The resulting cubes cover $-300$ to $+300$~\kms{} in $V_{\rm LSR}$.

We binned the spectra by a factor of two to improve the signal-to-noise ratio. The resulting velocity resolution is 2~\kms{}.
Because some residual baseline structure remained after gridding, we refitted and subtracted baselines using carefully selected line-free ranges.
Most spectra were fitted with a first-order polynomial, whereas those showing stronger baseline distortions were fitted with a third-order polynomial.
The conversion from $T_{\rm a}^*$ to main-beam temperature $T_{\rm MB}$ was made using the beam efficiencies $\eta_{\rm MB}$ listed in Table~\ref{tab:obs}.
The typical rms noise levels measured across the full survey area are also listed in Table~\ref{tab:obs}.
The effective angular resolution is $20\arcsec$, corresponding to 0.8~pc at 8.28~kpc.

Because the rest-frequency separation between the SiO and H$^{13}$CO$^+$ lines corresponds to only $\sim$320~\kms{}, the two lines inevitably overlap in velocity space over the broad velocity range of the CMZ (see the unmasked channel maps in Appendix~\ref{sec:refchanmaps}).
To avoid this line confusion, we constructed a masking template from the CS emission using the automated dual-threshold masking procedure implemented in \texttt{takefits} \citep{takekawa_takefits}.
Following a seed-and-grow strategy \citep{rosolowsky06}, we defined emission seeds as voxels above 4.5$\sigma$ in the smoothed CS cube and then expanded the mask to connected voxels above 3$\sigma$ in the unsmoothed CS cube.
We then applied this CS-based mask to the SiO and H$^{13}$CO$^+$ cubes.

The released survey products include calibrated FITS cubes, integrated-intensity maps, analysis masks, the SiO cloud catalog, and the cloud-integrated line-intensity table.
The calibrated FITS products are available in Zenodo at \dataset[doi:10.5281/zenodo.19588616]{https://doi.org/10.5281/zenodo.19588616}.

%% ============================================================

\section{Large-Scale Molecular Line Maps} \label{sec:results}
 \begin{figure*}[t!]
\figurenum{1}
\epsscale{1.15}
\plotone{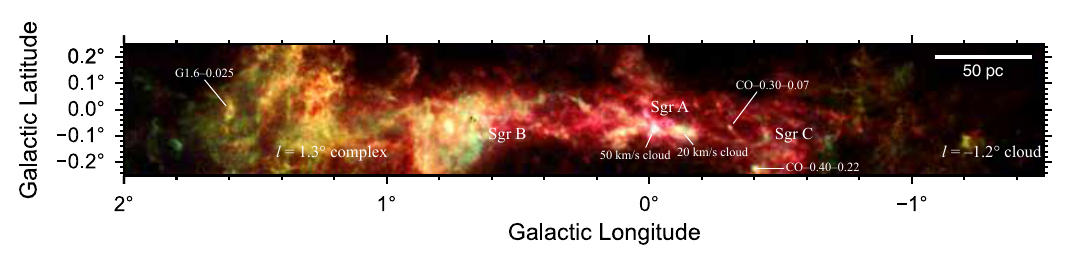}
\caption{Three-color composite map of the CMZ.
Red, green, and blue represent the integrated intensities of the HCN $J$=1--0, SiO $J$=2--1, and H$^{13}$CN $J$=1--0 lines, respectively.
The scale bar corresponds to 50~pc, assuming a Galactic center distance of 8.28~kpc. \label{fig:overview}}
\end{figure*}

\subsection{Integrated-intensity Maps} \label{sec:maps}

Figure~\ref{fig:overview} presents a three-color composite map of the CMZ in integrated intensities of the HCN $J$=1--0 (red), SiO $J$=2--1 (green), and H$^{13}$CN $J$=1--0 (blue) lines.
The map shows a large-scale east--west asymmetry in the relative strength of SiO emission.
Purple and magenta tones along the main dense gas ridge, most prominently toward Sgr~A, arise from bright HCN and H$^{13}$CN with relatively weaker SiO, consistent with dense gas in which HCN is relatively optically thick.
Yellow and green tones are particularly prominent in the Sgr~B complex, the eastern CMZ including the $l=1\fdg3$ complex, and the $l=-1\fdg2$ cloud, where dominant SiO emission suggests stronger shock activity.
White regions are most evident toward the 20 and 50~\kms{} clouds (M--0.13--0.08 and M--0.02--0.07) in the Sgr~A complex and parts of the Sgr~B complex, where dense gas and shocked gas are both prominent.
Several small yellowish to whitish clumps are also discernible, including HVCCs such as CO--0.40--0.22 \citep{oka16} and CO--0.30--0.07 \citep{tanaka15}.
The composite map shows that the SiO-traced shocked gas is distributed across the CMZ in both extended structures and compact clumps, rather than simply following the brightest dense gas structures.

\begin{figure*}[t!]
\figurenum{2}
\epsscale{1.15}
\plotone{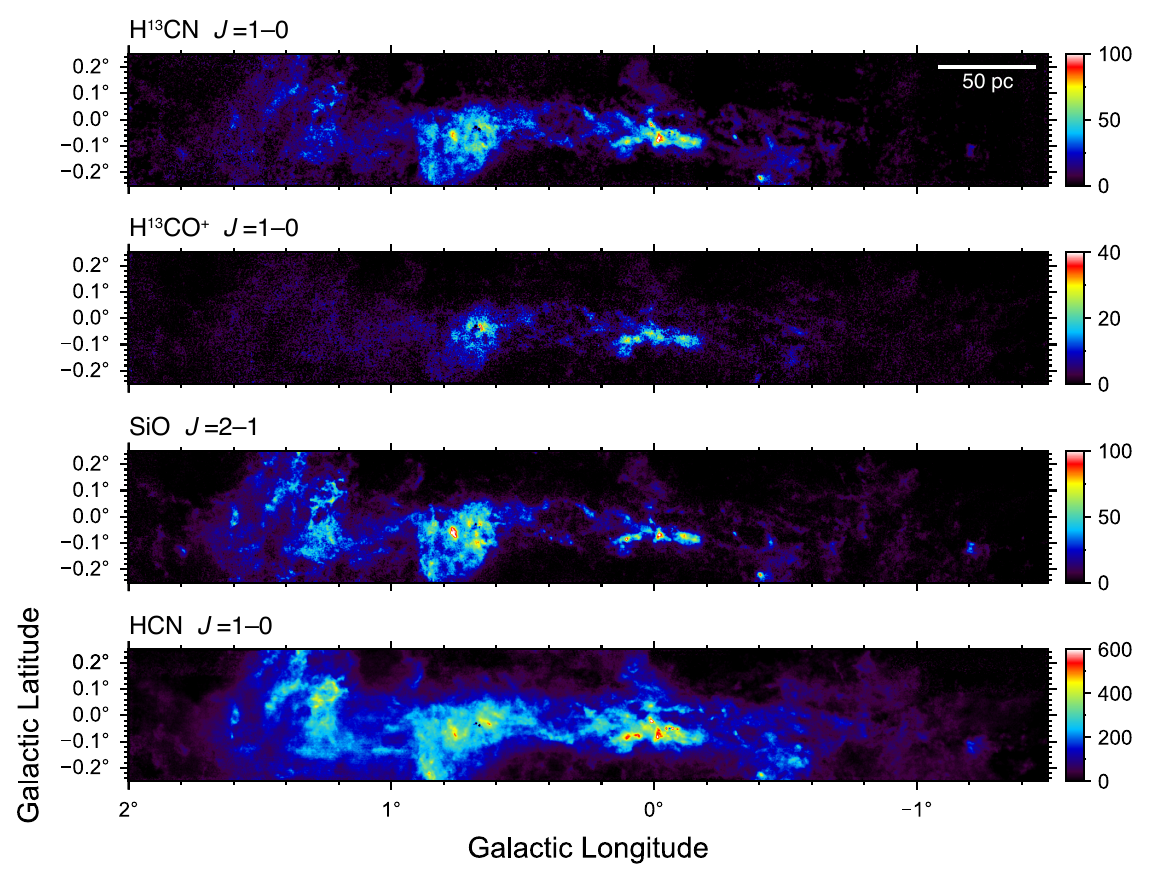}
\caption{Integrated intensity maps of H$^{13}$CN $J$=1--0, H$^{13}$CO$^+$ $J$=1--0, SiO $J$=2--1, and HCN $J$=1--0, from top to bottom. The color scale indicates the integrated intensity in units of K~\kms{}. The scale bar in the upper-right corner corresponds to 50~pc at a distance of 8.28~kpc. \label{fig:maps1}}
\end{figure*}

\begin{figure*}[t!]
\figurenum{3}
\epsscale{1.15}
\plotone{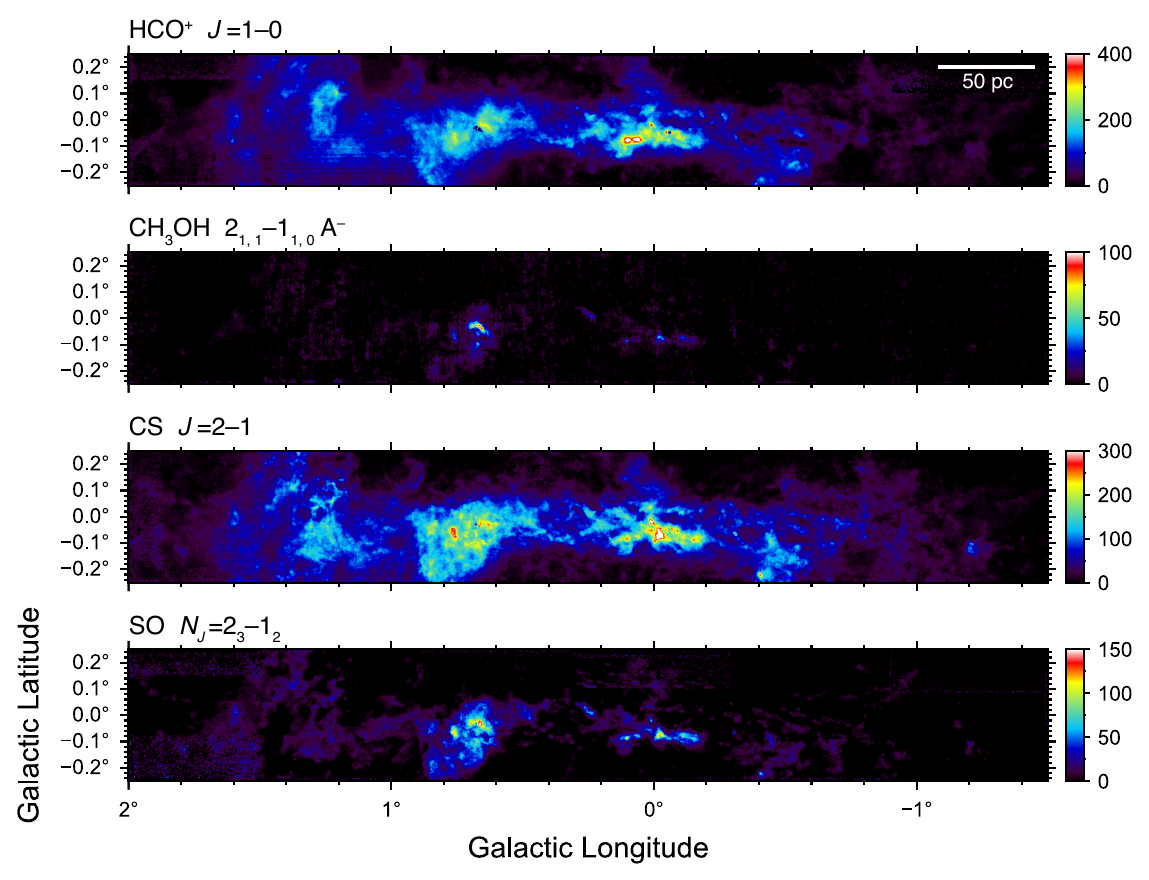}
\caption{Integrated intensity maps of HCO$^+$ $J$=1--0, CH$_3$OH $2_{1,1}$--$1_{1,0}$~A$^-$, CS $J$=2--1, and SO $N_J$=$2_3$--$1_2$, from top to bottom. The color scale indicates the integrated intensity in units of K~\kms{}. The scale bar in the upper-right corner corresponds to 50~pc at a distance of 8.28~kpc. \label{fig:maps2}}
\end{figure*}

Figures~\ref{fig:maps1} and \ref{fig:maps2} show the individual integrated-intensity maps of all eight observed molecular lines.
The large-scale spatial structures seen in the composite (Figure~\ref{fig:overview}) are recovered across both figures, and the individual maps further clarify the behavior of each tracer.
HCN, HCO$^+$, and CS show the most extended spatial distributions. Their continuous emission spans from $l \approx -1\fdg2$ to $+1\fdg6$.
Among the eight lines, HCN reaches the highest peak integrated intensity and H$^{13}$CO$^+$ the lowest, while CH$_3$OH shows the most localized distribution.
SO is concentrated toward the Sgr~A and Sgr~B complexes, though faint extended emission is also visible across the CMZ.

Because the Setup~2 lines (SO, HCN, and HCO$^+$) were observed only in the x-scan direction, residual baseline distortions that could not be fully removed produce stripe-like features in the maps. These distortions appeared irregularly in some specific spectrometer bands and their origin remains unclear.
These artifacts are most noticeable in the SO map at $b \gtrsim 0\fdg1$ for $-0\fdg3 \lesssim l \lesssim 0\fdg3$, and at both $b \gtrsim 0\fdg15$ and $b \lesssim -0\fdg07$ for $l \gtrsim 1\fdg5$.
Despite localized variations in noise level, the maps provide wide-field coverage with sufficient sensitivity to trace the large-scale molecular structure of the CMZ.

\subsection{Velocity-Channel Maps} \label{sec:chanmaps}
\begin{figure*}[p!]
\figurenum{4}
\centering
\includegraphics[height=0.78\textheight,keepaspectratio]{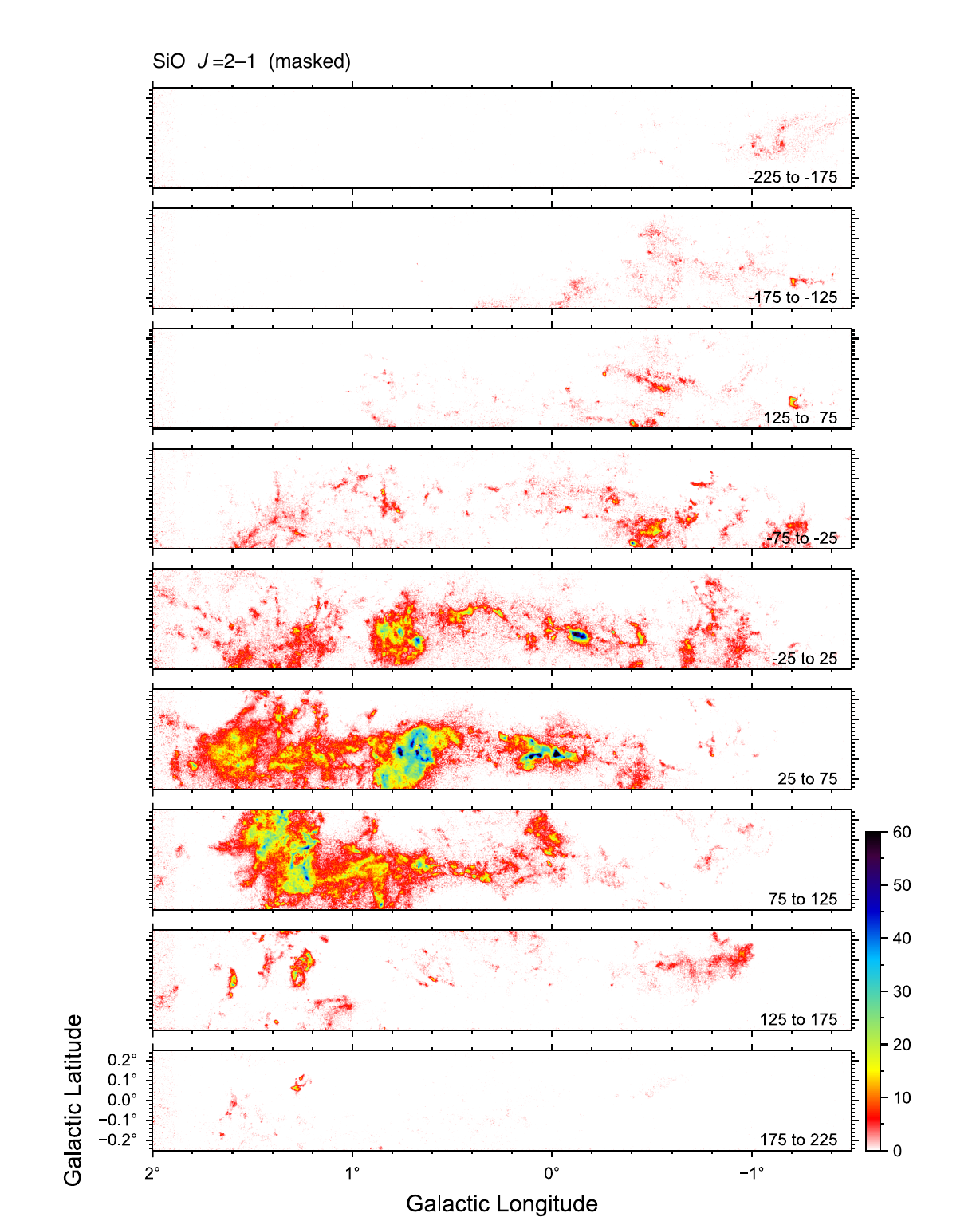}
 \caption{Velocity-channel maps of the SiO $J$=2--1 line.
Each panel shows the integrated intensity over a 50~\kms{} interval from $-225$ to $+225$~\kms{}.
The CS-based mask described in Section~\ref{sec:obs} has been applied to reduce mutual contamination with the nearby H$^{13}$CO$^+$ line.
  \label{fig:sio_chmap} }
\end{figure*}

\begin{figure*}[p!]
\figurenum{5}
\centering
\includegraphics[height=0.78\textheight,keepaspectratio]{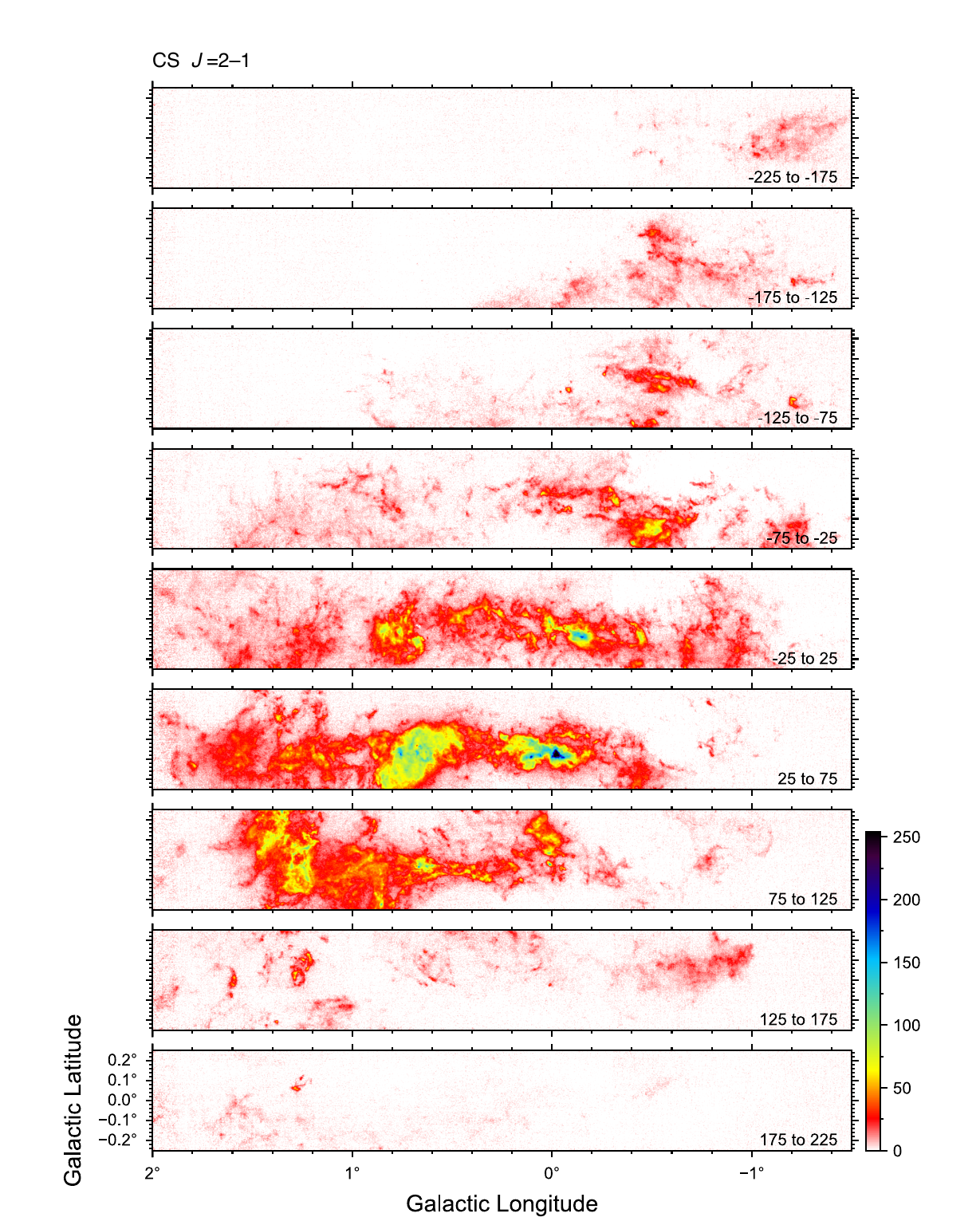}
 \caption{Same as Figure~\ref{fig:sio_chmap}, but for the CS $J$=2--1 line.
  \label{fig:cs_chmap_main} }
\end{figure*}

\begin{figure*}[p!]
\figurenum{6}
\centering
\includegraphics[height=0.78\textheight,keepaspectratio]{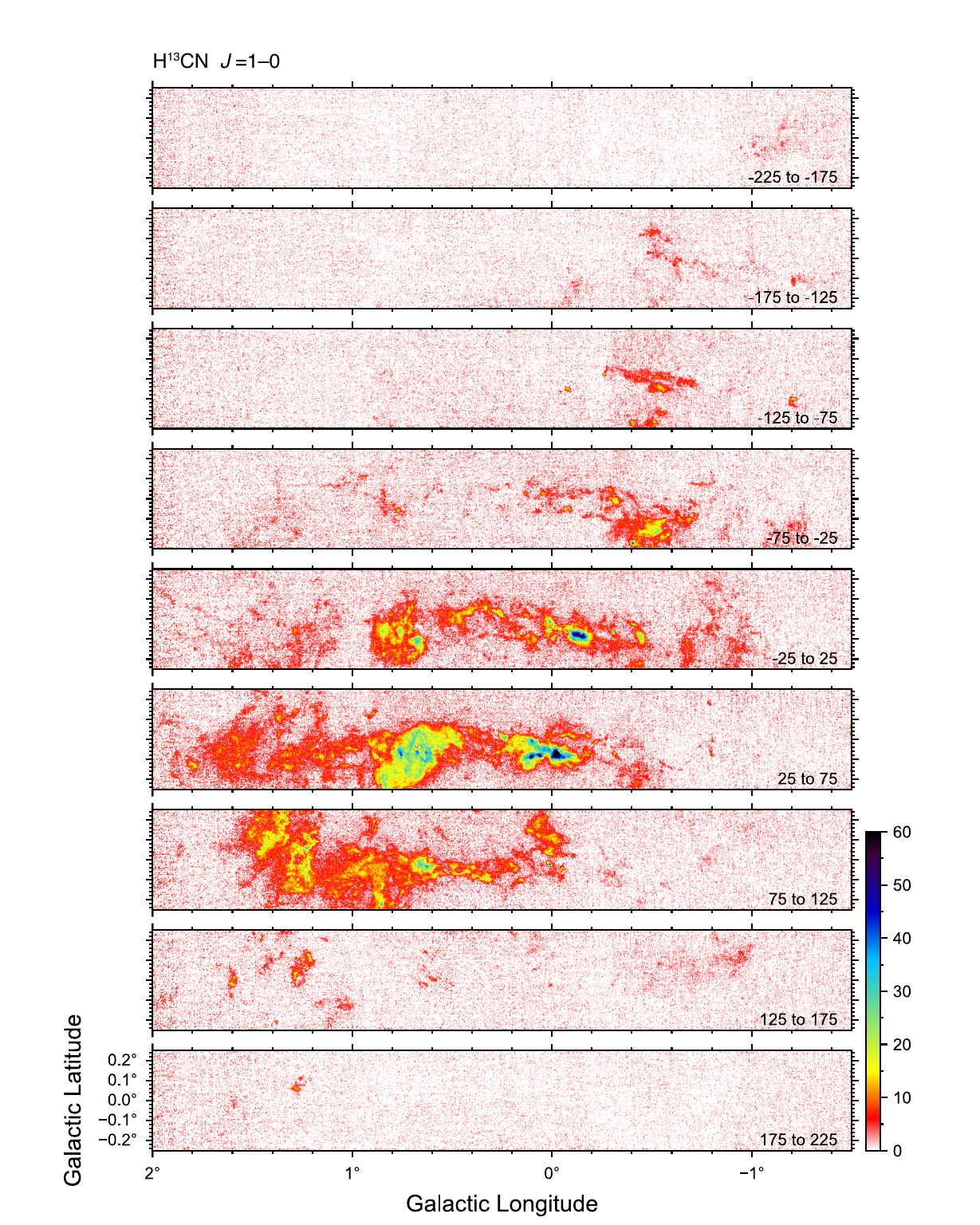}
\caption{Same as Figure~\ref{fig:sio_chmap}, but for the H$^{13}$CN $J$=1--0 line.
}\label{fig:h13cn_chmap_main}
\end{figure*}

Figures~\ref{fig:sio_chmap}--\ref{fig:h13cn_chmap_main} show representative velocity-channel maps of the SiO, CS, and H$^{13}$CN lines.
CS provides a bright dense-gas reference and traces the major CMZ complexes over a broad velocity range.
H$^{13}$CN recovers the same principal structures with lower brightness and reduced diffuse emission relative to HCN, supporting its use as an optically thinner reference line for the cloud-mass estimates.
SiO is widespread but more intermittent, appearing in clumps and filaments rather than simply following the brightest dense-gas emission.
This behavior is expected for a molecule enhanced by recent shocks and subsequent dust-grain processing.

The maps are integrated over nine 50~\kms{} intervals from $-225$ to $+225$~\kms{} and provide a velocity-resolved view of the molecular emission across the CMZ.
These maps reproduce the large-scale structures previously reported \citep[e.g.,][]{jones12} and reveal fainter extended emission across the CMZ.
Emission in these tracers is concentrated mainly between approximately $-175$ and $+175$~\kms{}.
The brightest and most extended structures appear between approximately $-75$ and $+125$~\kms{}, where the major CMZ cloud complexes dominate.
At positive longitudes, the high-velocity emission is prominent toward the $l=1\fdg3$ complex and the G1.6$-$0.025 cloud.
In the $+125$ to $+175$~\kms{} channels, an extended component appears to trail eastward from $l \sim -1\fdg0$ and is likely associated with the expanding molecular ring (EMR; \citealt{kaifu72, scoville72}).
The negative-velocity counterpart of the EMR is also visible, especially in CS at $l \lesssim 1\fdg0$ for $V_{\rm LSR}<-75$~\kms{}, and is faintly detected even in SiO and H$^{13}$CN.
In SiO, the $l=-1\fdg2$ cloud is one of the most prominent structures at $V_{\rm LSR}<-75$~\kms{} \citep{tsujimoto18}.
The velocity-resolved SiO morphology defines the parent data cube for the SCIMES cloud catalog analyzed in the following sections.

As an additional kinematic reference, Appendix~\ref{sec:refchanmaps} provides longitude--velocity maps averaged over five 0\fdg1-wide latitude bins spanning $-0\fdg25 \leq b \leq +0\fdg25$.
Channel maps of the other observed transitions, together with unmasked SiO and H$^{13}$CO$^+$ comparison maps, are presented there as well.
HCN and HCO$^+$ trace the extended optically thick dense-gas envelope, H$^{13}$CO$^+$ provides a weaker optically thinner comparison, and CH$_3$OH and SO highlight more localized chemically processed gas.
Detailed chemical interpretation and analyses of individual peculiar features are deferred to subsequent papers.

\clearpage
\section{SiO-emitting Cloud Catalog and SiO Line-ratio Trends} \label{sec:cloudcatalog}

We next quantify the SiO-emitting structures identified in the survey maps.
This section describes the cloud sample, the cloud-integrated line measurements and H$^{13}$CN-based physical quantities, and trends in SiO intensity ratios across the cloud sample.

\subsection{Identification of SiO-emitting Clouds}   \label{sec:cloudid}
\begin{figure*}[bhp]
  \figurenum{7}
 \epsscale{1.2}
\plotone{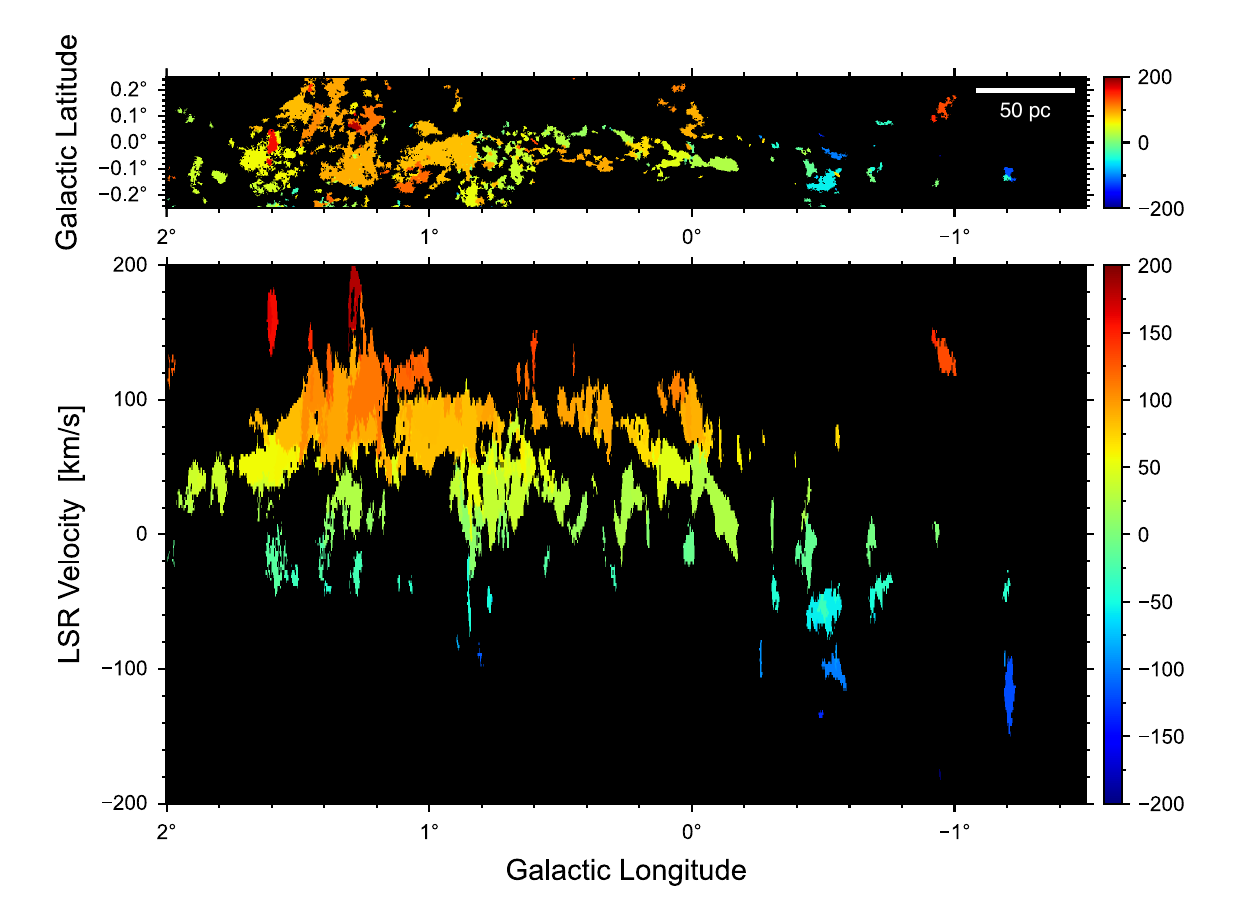}
\caption{Distribution of the SiO-emitting clouds identified by SCIMES in the $l$--$b$ plane (top) and $l$--$V$ plane (bottom). Individual clouds are color coded by their centroid velocity $V_{\rm cen}$.} \label{fig:clouds}
\end{figure*}

 To quantify the SiO structures in a uniform way across the CMZ, we compile a catalog of SiO-emitting clouds.
We use the same reference sample of SiO-emitting clouds identified from the SiO data cube with the Spectral Clustering for Interstellar Molecular Emission Segmentation (SCIMES) algorithm \citep{colombo15} as in \citet{takekawa24}.
We adopted the same dendrogram parameters as \citet{takekawa24}: \texttt{min\_value}=\texttt{min\_delta}=0.39~K ($3\sigma$) and \texttt{min\_npix}=34.
The catalog contains 258 clouds distributed across the CMZ.
The spatial and velocity distributions of the cataloged clouds are shown in Figure~\ref{fig:clouds}.

For each cloud, we measured the integrated intensities of all eight lines by summing the emission within the SiO-defined SCIMES masks.
The same three-dimensional mask is used for all lines of a given cloud.
We also measure the size parameter $S$, effective radius $R_{\rm eff}$, and velocity dispersion $\sigv{}$.
Here, the size parameter is $S = \sqrt{\sigma_{\rm maj}\sigma_{\rm min}}$, where $\sigma_{\rm maj}$ and $\sigma_{\rm min}$ are the intensity-weighted rms sizes along the major and minor axes of the cloud. The effective radius is $R_{\rm eff} = \sqrt{A / \pi}$, where $A$ is the projected area of the cloud.
The integrated intensities of all eight lines are listed in Appendix~\ref{sec:catalogproducts}.

\subsection{Derived Physical Properties} \label{sec:cloudprops}

We derive the cloud masses from the H$^{13}$CN line, whose emission broadly follows the dense gas structures traced by HCN, HCO$^+$, and CS across the CMZ.
We first assess whether H$^{13}$CN is a suitable column-density tracer by estimating its effective optical depth from the cloud-integrated HCN/H$^{13}$CN intensity ratio, adopting a $^{12}$C/$^{13}$C isotopic abundance ratio of 24 \citep{langer90}.
The derivation of the effective optical depth, excitation temperature, and column density is summarized in Appendix~\ref{sec:lteform}.
Figure~\ref{fig:cloud_tau_tex_hist} shows the histograms of the derived optical depth and excitation temperature.
Solving for the effective optical depth under a common excitation temperature gives a median $\tau(\mathrm{H^{13}CN}) \approx 0.19$, and about 90\% of the clouds have $\tau < 0.5$.
The H$^{13}$CN line is therefore optically thin in most of the sample and is suitable as a column-density tracer.

 \begin{figure*}[t!]
\figurenum{8}
\epsscale{1.1}
\plotone{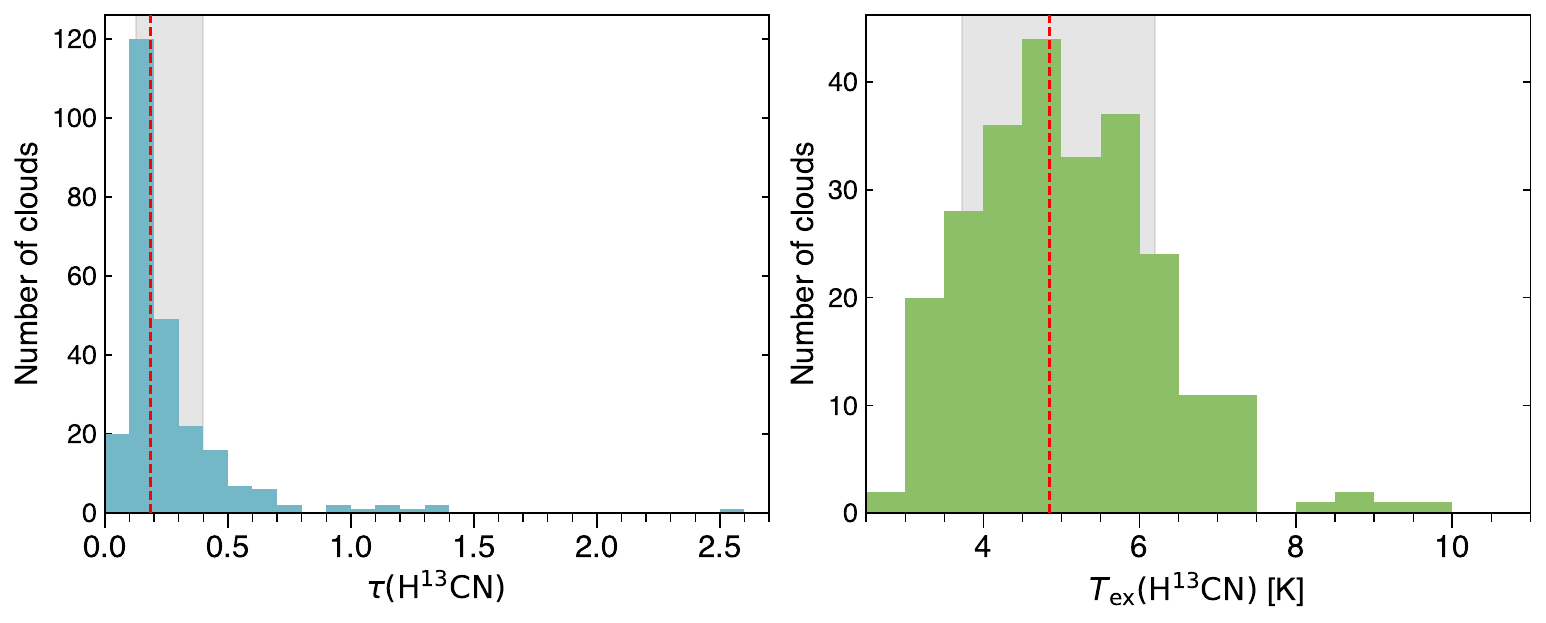}
\caption{Histograms of the effective H$^{13}$CN optical depth (left) and effective excitation temperature (right) for the cataloged clouds. The definitions and calculation of these quantities are given in Appendix~\ref{sec:lteform}.
The red dashed lines show the median values, and the shaded ranges indicate the 16th--84th percentiles.
\label{fig:cloud_tau_tex_hist}}
\end{figure*}

For the H$^{13}$CN column-density calculation, we adopt an excitation temperature of $T_{\rm ex} = 5$~K, consistent with the median value ($T_{\rm ex} =4.84$ K) inferred from the HCN/H$^{13}$CN ratio and the cloud-averaged H$^{13}$CN brightness temperature.
Under the assumptions of local thermodynamic equilibrium (LTE) and optically thin conditions, we convert the H$^{13}$CN intensity in each cloud mask to a column density using the standard formulation summarized in Appendix~\ref{sec:lteform}.
For each spatial pixel associated with a cloud, the H$^{13}$CN intensity is integrated only over the velocity channels belonging to that cloud.
Using a fractional abundance $X(\mathrm{H^{13}CN}) \equiv [\mathrm{H^{13}CN}]/[\mathrm{H_2}] = 4 \times 10^{-10}$ \citep{goicoechea18}, we then convert $N(\mathrm{H^{13}CN})$ to $N(\mathrm{H_2})$ and derive the cloud mass $M_{\rm H_2}$.
For a cloud $c$, the mass is estimated as
\begin{equation}
M_{{\rm H_2}, c} =
\frac{\mu m_{\rm H} D^2 \Omega_{\rm pix}}{X(\mathrm{H^{13}CN})}
\sum_{p\in c} N_p(\mathrm{H^{13}CN}),
\end{equation}
where $N_p(\mathrm{H^{13}CN})$ is the H$^{13}$CN column density in spatial pixel $p$ within the cloud mask, $\mu=2.8$ is the mean molecular weight per H$_2$ molecule, $m_{\rm H}$ is the hydrogen atom mass, $D=8.28$~kpc, and $\Omega_{\rm pix}=(7.5\text{ arcsec})^2$ is the solid angle of one spatial pixel.
This procedure gives a total H$^{13}$CN-based mass of $4.0 \times 10^6~M_\odot$ for the SCIMES cloud sample.
Applying the same LTE assumptions, the total gas mass traced by H$^{13}$CN in the full survey area is $\sim 2.8 \times 10^7~M_\odot$.
Although H$^{13}$CN does not trace the same gas component as CO or dust continuum, this full-area mass is comparable to previous CMZ molecular gas mass estimates of a few $10^7~M_\odot$ \citep[e.g.,][]{morris96, dahmen98, oka98, molinari11}.
This similarity is consistent with the large dense-gas fraction in the CMZ and suggests that the adopted H$^{13}$CN abundance gives a reasonable mass scale for the present analysis.
The summed mass of the SCIMES SiO-emitting clouds ($\sim 4.0 \times 10^6~M_\odot$) is therefore about 15\% of the total H$^{13}$CN-derived gas mass over the survey area ($\sim 2.8 \times 10^7~M_\odot$).
This indicates that the SiO-bright shocked component is significant but does not dominate the dense molecular gas in the CMZ.

The cloud-averaged H$_2$ number density is given by $\nbar{} = M_{\rm H_2} / (\mu m_{\rm H} V)$, where $V = (4\pi/3)R_{\rm eff}^3$ is the cloud volume assuming spherical geometry.
This is a volume-averaged density over each SCIMES cloud mask and should not be interpreted as the local density of the densest substructure within the cloud.
The dynamical time is defined as $\tdyn{} = S / \sigv{}$.
For clouds with valid H$^{13}$CN-based masses, the median values are about $1.9 \times 10^3$~\cmcube{} for $\nbar{}$ and $1.5 \times 10^5$~yr for $\tdyn{}$.
The corresponding 10th--90th percentile ranges are $7 \times 10^2$--$5 \times 10^3$~\cmcube{} and $8 \times 10^4$--$3 \times 10^5$~yr, respectively.
Table~\ref{tab:catalog} lists the positions, velocities, sizes, velocity dispersions, H$^{13}$CN-based masses, cloud-averaged densities, and dynamical times of the SiO-emitting clouds.

The median $\nbar{}$ of $\sim 1.9 \times 10^3$~\cmcube{} is lower than the mean densities of order $10^4$~\cmcube{} reported for CMZ molecular gas \citep[e.g.,][]{kruijssen14}.
This density difference is expected because $\nbar{}$ is a volume average over the SCIMES mask.
If the dense H$^{13}$CN/SiO-emitting substructure occupies a volume filling factor $f_V<1$, its characteristic local density is higher by a factor of $\sim f_V^{-1}$.
At the individual-cloud scale, we compare two well-studied regions that can be clearly matched to individual cataloged SiO clouds and have published whole-cloud mass estimates.
G0.253+0.016, commonly known as the Brick, is matched to SiO cloud ID~59 at $(l,b,V)=(0\fdg254,0\fdg020,24.0~{\rm km~s^{-1}})$.
Its H$^{13}$CN-based mass is $6.8\times10^4~M_\odot$, comparable to the $1.3\times10^5~M_\odot$ mass of the full cloud reported by \citet{longmore12}.
The Sgr~A 20/50~\kms{} cloud complex is represented by a single dominant SiO structure, ID~37, at $(l,b,V)=(-0\fdg085,-0\fdg076,25.5~{\rm km~s^{-1}})$, whose H$^{13}$CN-based mass is $4.6\times10^5~M_\odot$.
This is comparable to the sum of the CO-derived H$_2$ masses of the 20 and 50~\kms{} clouds, $2.9\times10^5$ and $1.9\times10^5~M_\odot$, respectively, reported by \citet{yang26}.
More extensive cloud-by-cloud comparisons will be presented in a forthcoming paper.

\begin{deluxetable*}{lccccccrrc}[htb]
\tabletypesize{\scriptsize}
\tablenum{2}
\tablecaption{Catalog of SiO-Emitting Clouds \label{tab:catalog}}
\tablewidth{0pt}
\tablehead{
\colhead{ID} & \colhead{$l$} & \colhead{$b$} & \colhead{$V_{\rm LSR}$} & \colhead{$S$} & \colhead{$R_{\rm eff}$} & \colhead{$\sigv{}$} & \colhead{$M_{\rm H_2}$} & \colhead{$\nbar{}$} & \colhead{$\tdyn{}$} \\
\colhead{} & \colhead{(deg)} & \colhead{(deg)} & \colhead{(\kms{})} & \colhead{(pc)} & \colhead{(pc)} & \colhead{(\kms{})} & \colhead{($10^2~M_\odot$)} & \colhead{($10^3$~\cmcube{})} & \colhead{($10^5$~yr)}
}
\startdata
1 & $-1.20$ & $-0.11$ & $-120$ & $1.44$ & $3.31$ & $10.8$ & $234.2$ & $2.5$ & $1.3$ \\
2 & $-1.19$ & $-0.13$ & $-41$ & $0.86$ & $1.95$ & $5.4$ & $15.7$ & $0.81$ & $1.6$ \\
3 & $-1.19$ & $-0.13$ & $-93$ & $0.38$ & $0.95$ & $3.5$ & $3.6$ & $1.6$ & $1.1$ \\
4 & $-0.96$ & $+0.14$ & $132$ & $2.66$ & $4.40$ & $6.3$ & $86.5$ & $0.39$ & $4.1$ \\
5 & $-0.94$ & $-0.05$ & $-178$ & $0.26$ & $0.66$ & $2.3$ & $2.8$ & $3.8$ & $1.1$ \\
$\vdots$ & & & & & & & & & \\
\enddata
  \tablecomments{Definitions of the derived quantities are given in Sections~\ref{sec:cloudid} and \ref{sec:cloudprops}.
  (This table is available in its entirety in machine-readable form in the online article.)}
 \end{deluxetable*}

\subsection{ SiO Line-ratio Trends} \label{sec:ratios}

\begin{figure*}[t!]
  \figurenum{9}
\plotone{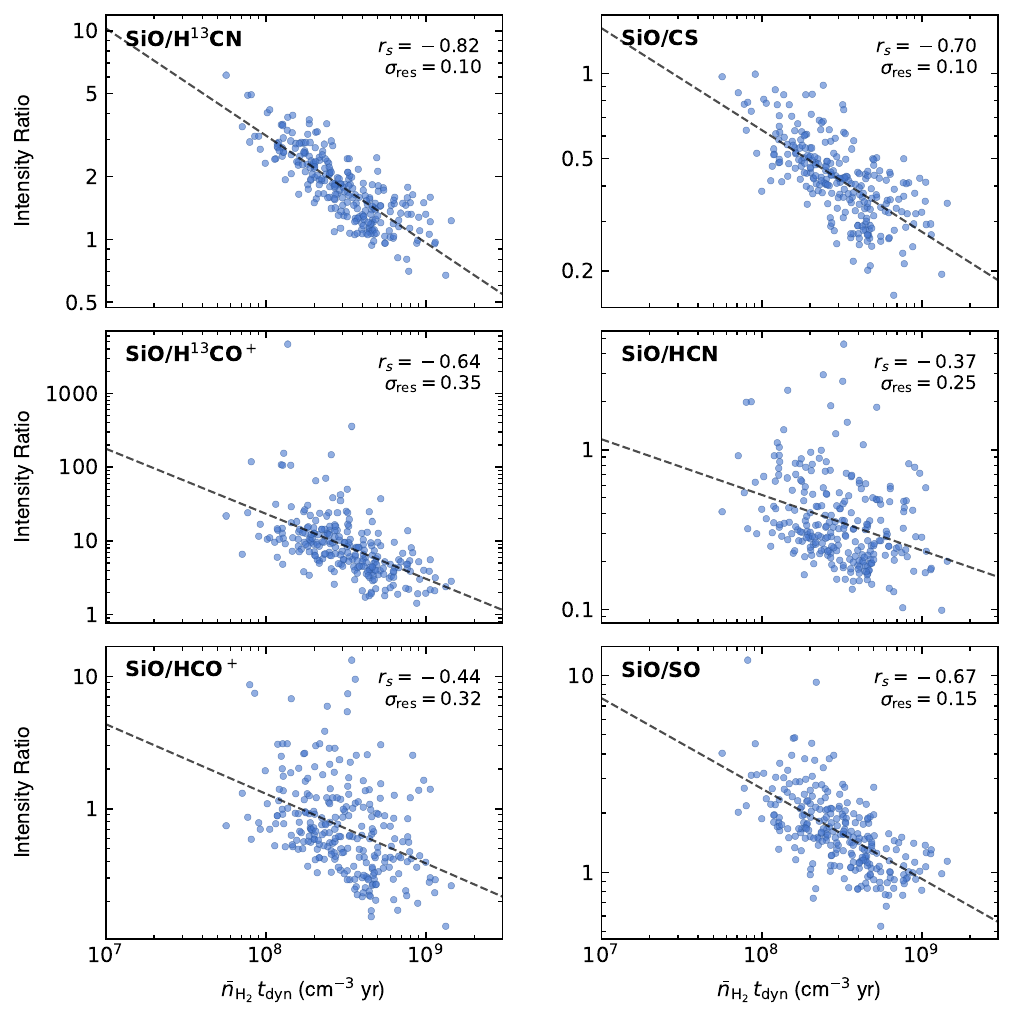}
 \caption{SiO intensity ratios as a function of $\ntdyn{}$ for six reference lines, SiO/H$^{13}$CN, SiO/CS, SiO/H$^{13}$CO$^+$, SiO/HCN, SiO/HCO$^+$, and SiO/SO.
The dashed line shows a log--log linear fit used as a visual guide and to estimate the residual scatter.
The Spearman coefficient $r_s$ and the residual scatter $\sigma_{\rm res}$ (in dex) are shown in each panel.
The analysis uses 255 clouds after excluding three clouds with unreliable H$^{13}$CN-based physical quantities.
The SiO/H$^{13}$CO$^+$ panel contains 243 clouds after excluding nonpositive H$^{13}$CO$^+$ intensities. \label{fig:main}}
\end{figure*}

\citet{takekawa24} found a parabolic-like spatial trend in the SiO line ratios across the CMZ. The ratios are lower near the Galactic nucleus and increase toward the edges (see their Figure~2).
This spatial trend may reflect variations in the postshock SiO depletion stage across the CMZ \citep{takekawa24}. 
To investigate how this spatial variation relates to cloud physical properties, we examine the correlations between the cloud-integrated SiO ratios and the physical properties derived in Section~\ref{sec:cloudprops}.
We use the density--time product $\ntdyn{}$ as a possible proxy for the progress of postshock SiO depletion.
After a shock, gas-phase SiO can be depleted by adsorption onto dust grains on a timescale inversely proportional to the gas density ($\tau_{\rm dep} \propto n^{-1}$; e.g., \citealt{martinpintado92, bergin98}).
Using $\nbar{}$ as a cloud-scale proxy for this density dependence and $\tdyn{}$ as a characteristic timescale for postshock evolution gives $\tdyn{}/\tau_{\rm dep} \propto \ntdyn{}$.
Under these assumptions, clouds with larger $\ntdyn{}$ would be expected to be at more advanced depletion stages and exhibit lower SiO ratios.
Figure~\ref{fig:main} plots the SiO intensity ratios relative to the other observed lines against the density--time product $\ntdyn{}$.
CH$_3$OH is excluded from this comparison because of its weak emission in many clouds.
Each ratio is calculated from the cloud-integrated intensities within the SCIMES masks.
Three clouds at the eastern survey boundary with unreliable H$^{13}$CN-based physical quantities are excluded, leaving 255 clouds.
The SiO/H$^{13}$CO$^+$ ratio additionally excludes 12 clouds with nonpositive H$^{13}$CO$^+$ integrated intensities, leaving 243 clouds for that ratio.

All six panels show that clouds with larger $\ntdyn{}$ have systematically lower SiO ratios.
The tightest correlations are seen for SiO/H$^{13}$CN and SiO/CS.
Their Spearman rank coefficients are $r_s = -0.82$ and $r_s = -0.70$.
 The residual scatter around log--log linear fits is 0.10~dex for both ratios.
The remaining four ratios show the same tendency, but with larger scatter.
Among them, SiO/SO shows the clearest negative trend ($r_s = -0.67$).
The SiO/H$^{13}$CO$^+$ ratio also follows this tendency ($r_s = -0.64$).
Its larger scatter ($\sigma_{\rm res} = 0.35$) is likely related to the weak H$^{13}$CO$^+$ intensities in many clouds.
The SiO/HCN and SiO/HCO$^+$ panels are more dispersed ($r_s = -0.37$ and $\sigma_{\rm res} = 0.25$ for SiO/HCN, and $r_s = -0.44$ and $\sigma_{\rm res} = 0.32$ for SiO/HCO$^+$).

Table~\ref{tab:spearman} summarizes the Spearman coefficients and the residual scatter $\sigma_{\rm res}$ from power-law fits for the correlations with $\ntdyn{}$, $\nbar{}$, and $\tdyn{}$.
For all six ratios, the correlation with $\ntdyn{}$ is tighter than those with either $\nbar{}$ or $\tdyn{}$ alone.
Figure~\ref{fig:ratio_n} shows the same six ratios plotted against $\nbar{}$ alone for comparison.
For most ratios, the values of $|r_s(\ntdyn{})|$ are larger by 0.1--0.2 than those of $|r_s(\nbar{})|$.
The correlations with $\tdyn{}$ alone remain weak for all six ratios ($|r_s(\tdyn{})| \leq 0.16$).
The residual scatter $\sigma_{\rm res}$ is consistently smaller for $\ntdyn{}$ than for $\nbar{}$.

  \begin{deluxetable*}{lccccc}[htb]
 \tabletypesize{\scriptsize}
\tablenum{3}
\tablecaption{Correlation Statistics for SiO Ratios \label{tab:spearman}}
\tablewidth{0pt}
  \tablehead{
\colhead{Ratio} & \colhead{$r_s(\ntdyn{})$} & \colhead{$\sigma_{\rm res}(\ntdyn{})$} & \colhead{$r_s(\nbar{})$} & \colhead{$r_s(\tdyn{})$} & \colhead{$\sigma_{\rm res}(\nbar{})$} \\
\colhead{} & \colhead{} & \colhead{(dex)} & \colhead{} & \colhead{} & \colhead{(dex)}
}
 \startdata
SiO/H$^{13}$CN &  $-0.82$ &  $0.10$ &  $-0.71$ & $+0.01$ &  $0.12$ \\
SiO/CS &  $-0.70$ &  $0.10$ &  $-0.58$ & $-0.03$ &  $0.12$ \\
SiO/H$^{13}$CO$^+$ &  $-0.64$ &  $0.35$ &  $-0.43$ & $-0.16$ &  $0.39$ \\
SiO/HCN &  $-0.37$ &  $0.25$ &  $-0.34$ & $+0.04$ &  $0.26$ \\
SiO/HCO$^+$ &  $-0.44$ &  $0.32$ &  $-0.35$ & $-0.04$ &  $0.33$ \\
SiO/SO &  $-0.67$ &  $0.15$ &  $-0.50$ & $-0.12$ &  $0.18$ \\
\enddata
\tablecomments{The residual scatter $\sigma_{\rm res}$ is measured around a linear fit in the log--log plane. The H$^{13}$CO$^+$ row uses 243 clouds after excluding nonpositive H$^{13}$CO$^+$ intensities. All other rows use 255 clouds.}
 \end{deluxetable*}

Because $\nbar{}$ and $\ntdyn{}$ depend on the H$^{13}$CN-based mass scale, the horizontal scale of Figure~\ref{fig:main} remains subject to excitation and abundance uncertainties.
The cloud-integrated HCN/H$^{13}$CN ratios indicate that H$^{13}$CN is optically thin over most of the cataloged clouds (Section~\ref{sec:cloudprops}), so optical-depth saturation should not dominate the mass scale.
A uniform multiplicative error in the H$^{13}$CN-to-H$_2$ conversion would shift $\nbar{}$ and $\ntdyn{}$ horizontally but would not by itself generate a rank-order correlation.

The same behavior across all six SiO ratios indicates that the observed correlations are likely driven by variations intrinsic to SiO.
Their systematic decline with increasing $\ntdyn{}$ is qualitatively consistent with subsequent depletion of shock-enhanced SiO in denser gas ($\tau_{\rm dep} \propto n^{-1}$; e.g., \citealt{martinpintado92}).
This interpretation is further supported by the tighter correlations with $\ntdyn{}$ than with either $\nbar{}$ or $\tdyn{}$ alone.

\begin{figure*}[t!]
\figurenum{10}
\epsscale{1.15}
\plotone{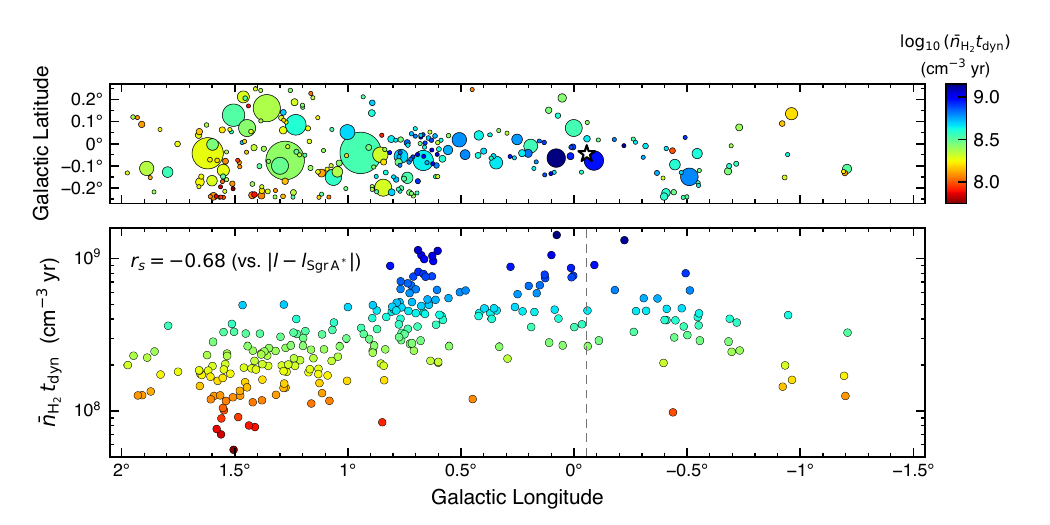}
\caption{Spatial distribution and longitude dependence of $\ntdyn{}$ for the cataloged SiO-emitting clouds.
The upper panel shows the spatial distribution of the SCIMES clouds.
The circle radii represent $R_{\rm eff}$, and the colors indicate $\log_{10}[\ntdyn{}/({\rm cm^{-3}\,yr})]$.
The white star marks Sgr~A$^*$.
The lower panel shows $\ntdyn{}$ as a function of Galactic longitude for the same clouds, using the same color scale as the upper panel.
The vertical dashed line marks the longitude of Sgr~A$^*$ ($l = -0\fdg056$), and the annotated Spearman coefficient is computed against the absolute longitude offset from Sgr~A$^*$, $|l - l_{\rm Sgr\,A^*}|$.
The cloud sample is the same as in Figure~\ref{fig:main} (255 clouds).\label{fig:nt_lon}}
\end{figure*}

Figure~\ref{fig:nt_lon} shows the spatial distribution of $\ntdyn{}$ across the CMZ.
The product $\ntdyn{}$ is anticorrelated with the absolute longitude offset from Sgr~A$^*$ ($r_s=-0.68$), with larger values toward central longitudes.
This trend is consistent with the position-based SiO-ratio trend reported by \citet{takekawa24}, in which lower SiO ratios are found toward the Galactic nucleus.

%% ============================================================
\section{Summary} \label{sec:summary}

We have presented molecular line maps of the CMZ obtained with the Nobeyama 45~m telescope.
The observations cover a $3\fdg5 \times 0\fdg5$ region at 20\arcsec\ resolution in eight molecular lines, including SiO $J$=2--1 and several tracers of dense gas.
The main results are summarized as follows.

\begin{enumerate}
\item Calibrated FITS cubes and integrated-intensity maps are released for all eight observed molecular lines.

\item Integrated intensity maps and channel maps characterize the distribution and kinematics of dense and shocked molecular gas across the CMZ.

\item A catalog of 258 clouds identified in SiO emission is compiled using the SCIMES algorithm.
For each cloud, the integrated intensities of all eight lines are tabulated using the SiO cloud masks.

\item The HCN/H$^{13}$CN ratios indicate that H$^{13}$CN is optically thin over most of the cataloged clouds.
Averaged molecular hydrogen densities $\nbar{}$ are derived from the masses estimated from H$^{13}$CN and volumes defined by the SiO cloud masks.
Dynamical times $\tdyn{}$ are estimated from the sizes and velocity dispersions of the SiO clouds.

\item SiO intensity ratios are compared with $\nbar{}$, $\tdyn{}$, and $\ntdyn{}$.
The ratios decline systematically with increasing $\ntdyn{}$.
The tightest correlations are found for SiO/H$^{13}$CN ($r_s=-0.82$) and SiO/CS ($r_s=-0.70$).
This dependence on the density--time product may trace depletion after shocks, as enhanced SiO in the gas phase depletes back onto dust grains.
\end{enumerate}

\begin{acknowledgments}
We thank the anonymous referee for constructive comments that improved the manuscript.
We are grateful to the staff of the Nobeyama Radio Observatory (NRO) for their outstanding support during the 45~m telescope observations.
The NRO is a branch of the National Astronomical Observatory of Japan (NAOJ), National Institutes of Natural Sciences.
This study was supported by JSPS KAKENHI grant Nos. JP19K14768, JP24K17091, and JP20H00178.
\end{acknowledgments}

\facility{No:45m}

\Needspace*{6\baselineskip}\software{
NOSTAR \citep{sawada08},
takefits \citep{takekawa_takefits},
Astropy \citep{astropy2013,astropy2018,astropy2022},
Matplotlib \citep{hunter2007},
NumPy \citep{vanderwalt2011,harris2020},
SciPy \citep{virtanen2020},
SCIMES \citep{colombo15}
}

%% ============================================================
%% Appendix
%% ============================================================

\onecolumngrid
\restartappendixnumbering
\appendix

\section{Correction for Off-Position Emission in the 2019 CS Observations} \label{sec:offpos}

\begin{figure}[ht!]
\figurenum{A1}
\plotone{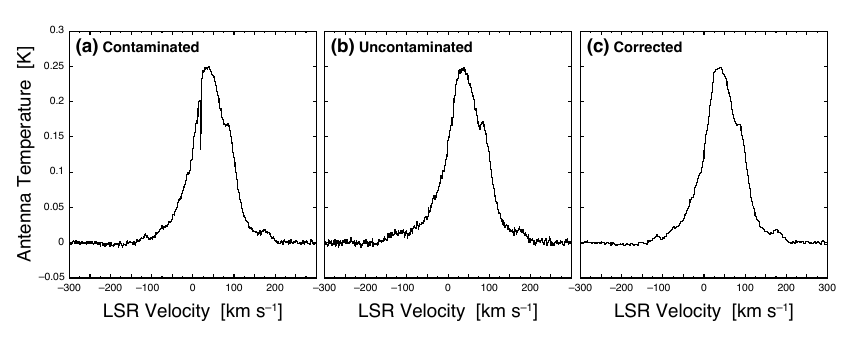}
\caption{(a) Average CS spectrum within $l = -0\fdg3$ to $+0\fdg9$ affected by the emission contamination from the off position at $(l, b) = (0\fdg0, -0\fdg5)$. The spurious absorption feature is visible at $V_{\rm LSR} \simeq 19$~\kms{}. (b) Average CS spectrum obtained using the uncontaminated off position at $b = -0\fdg75$. (c) Average CS spectrum after the correction procedure described in the text. \label{fig:offpos}}
\end{figure}

During the 2019 observations, spurious absorption lines centered at $V_{\rm LSR} \simeq 19$~\kms{} with a width of approximately 2~\kms{} were present in the CS spectra within $l = -0\fdg3$ to $+0\fdg9$ (Figure~\ref{fig:offpos}(a)).
These features were attributed to emission contamination from the off position at $(l, b) = (0\fdg0, -0\fdg5)$.

To mitigate this contamination, off positions at $b = -0\fdg75$ were adopted for observations conducted after 2020.
The spurious absorption lines in the affected 2019 data were corrected as follows.
Because the spurious feature showed little variation in depth and shape among the affected spectra, we treated it as a common contamination profile.
We first computed the average spectrum for each set of contaminated and uncontaminated spectra within $l = -0\fdg3$ to $+0\fdg9$.
The spurious line was isolated by subtracting the average uncontaminated spectrum (Figure~\ref{fig:offpos}(b)) from the average contaminated spectrum (Figure~\ref{fig:offpos}(a)).
The isolated spurious line was subsequently subtracted from each contaminated spectrum.
The average profile of the corrected spectra (Figure~\ref{fig:offpos}(c)) demonstrates that the off-position contamination has been effectively eliminated.

\section{Supplementary Velocity-Channel and Longitude--Velocity Maps} \label{sec:refchanmaps}

\begin{figure}[p!]
\figurenum{B1}
\centering
\includegraphics[height=0.78\textheight,keepaspectratio]{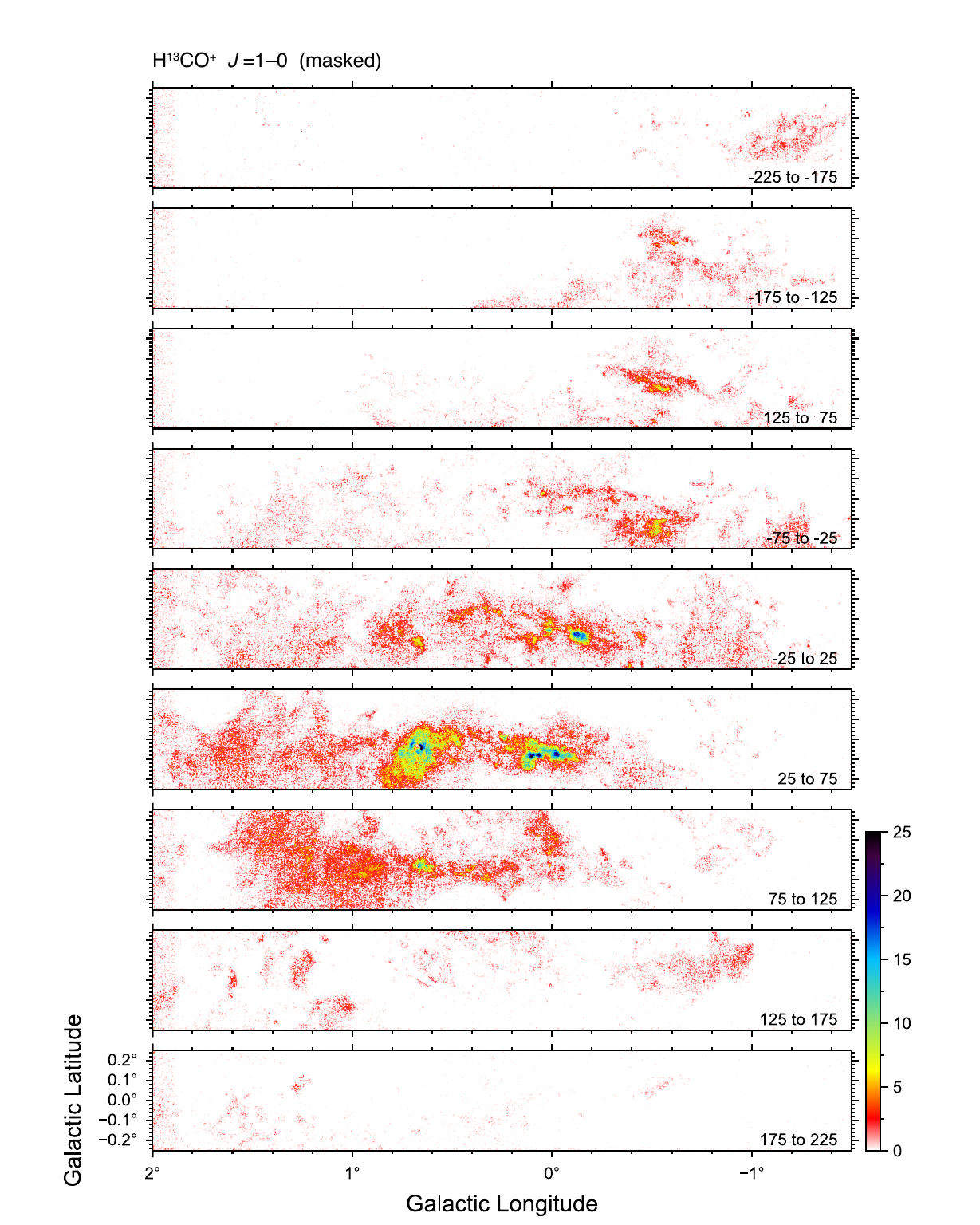}
\caption{Velocity-channel maps of the masked H$^{13}$CO$^+$ $J$=1--0 cube.
Each panel shows the integrated intensity over a 50~\kms{} interval from $-225$ to $+225$~\kms{}.
The CS-based mask described in Section~\ref{sec:obs} has been applied to reduce mutual contamination with the nearby SiO line.
\label{fig:h13cop_chmap}}
\end{figure}

\begin{figure}[p!]
\figurenum{B2}
\centering
\includegraphics[height=0.78\textheight,keepaspectratio]{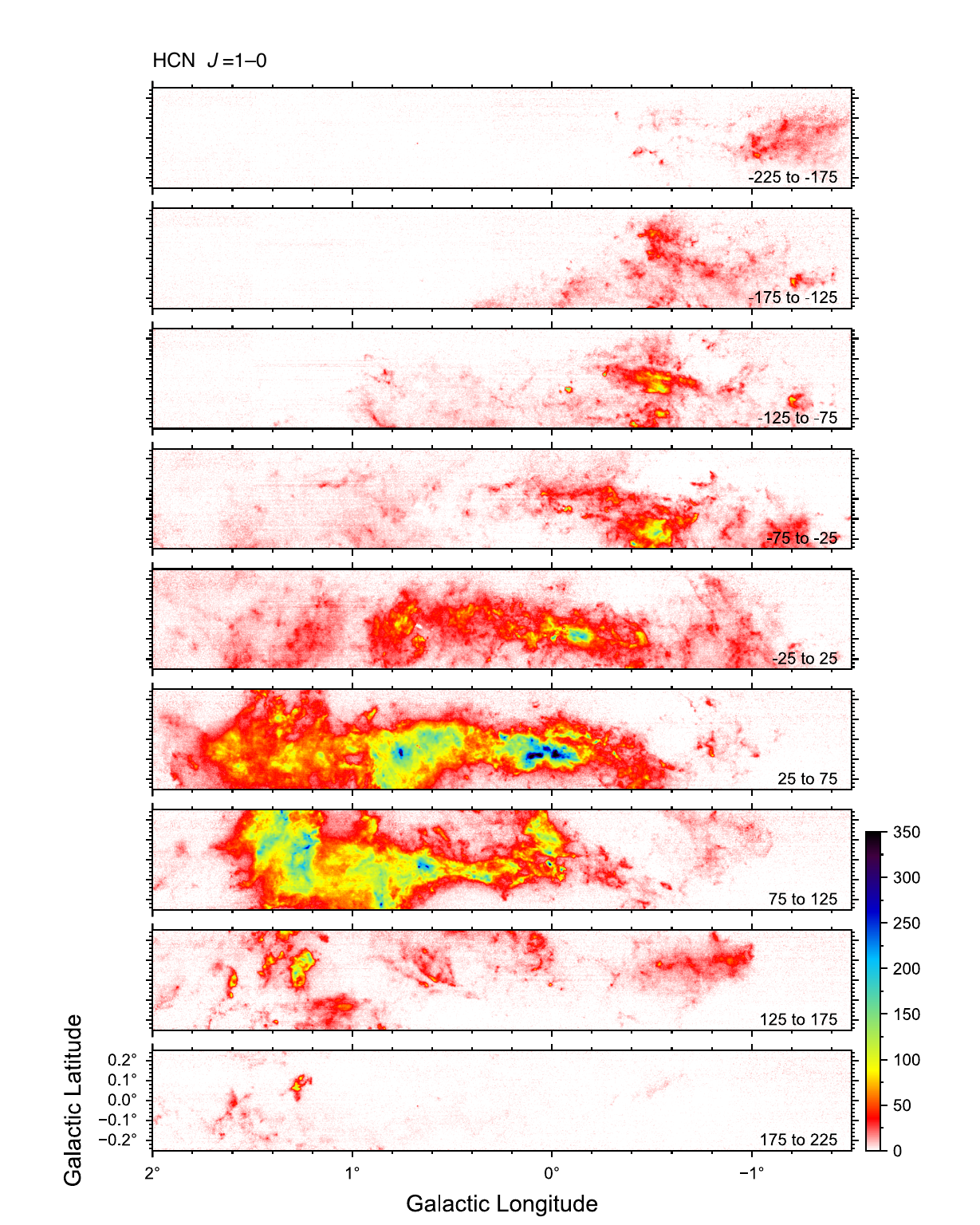}
\caption{Same as Figure~\ref{fig:h13cop_chmap}, but for the HCN $J$=1--0 line.
\label{fig:hcn_chmap}}
\end{figure}

\begin{figure}[p!]
\figurenum{B3}
\centering
\includegraphics[height=0.78\textheight,keepaspectratio]{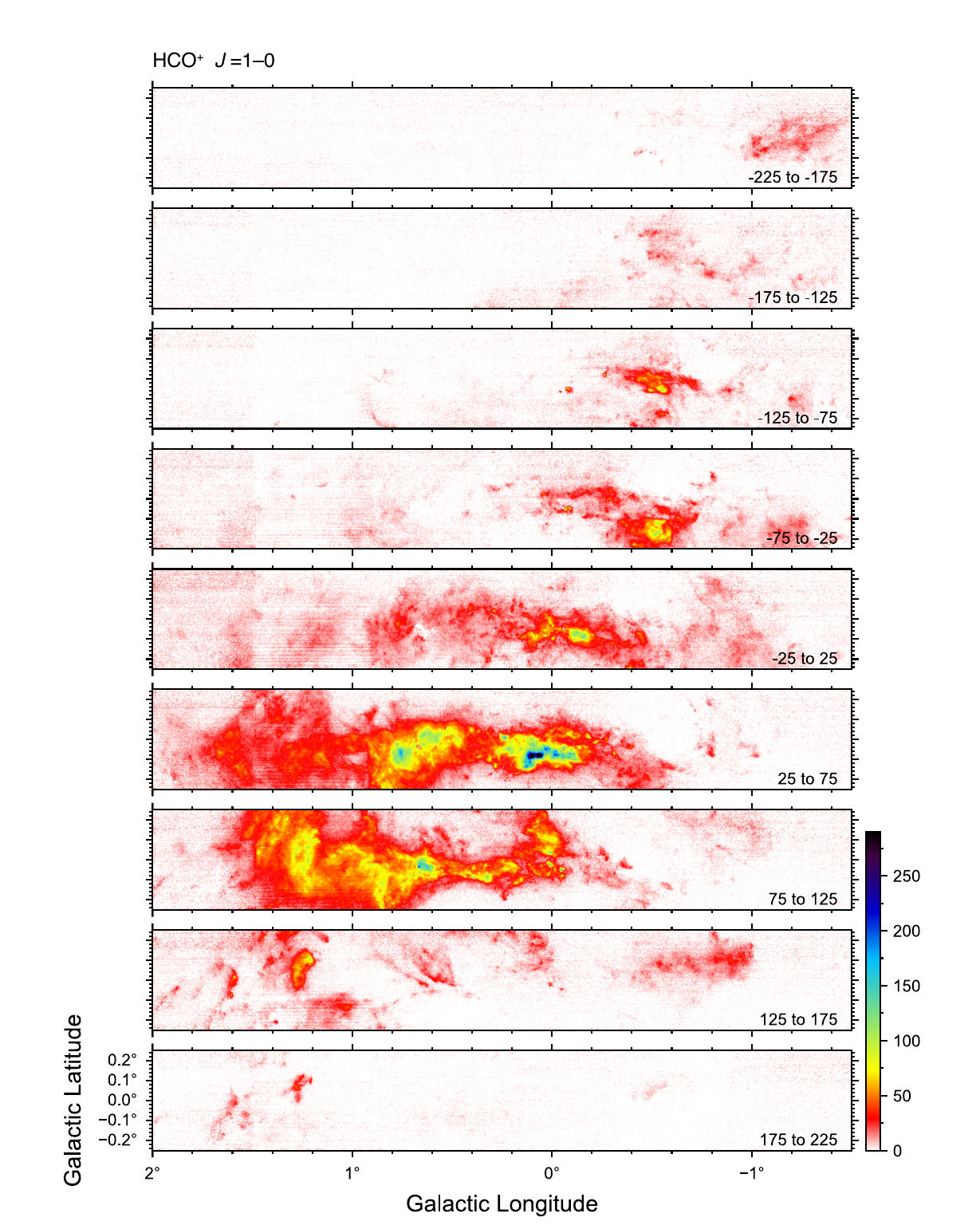}
\caption{Same as Figure~\ref{fig:h13cop_chmap}, but for the HCO$^+$ $J$=1--0 line.
\label{fig:hcop_chmap}}
\end{figure}

\begin{figure}[p!]
\figurenum{B4}
\centering
\includegraphics[height=0.78\textheight,keepaspectratio]{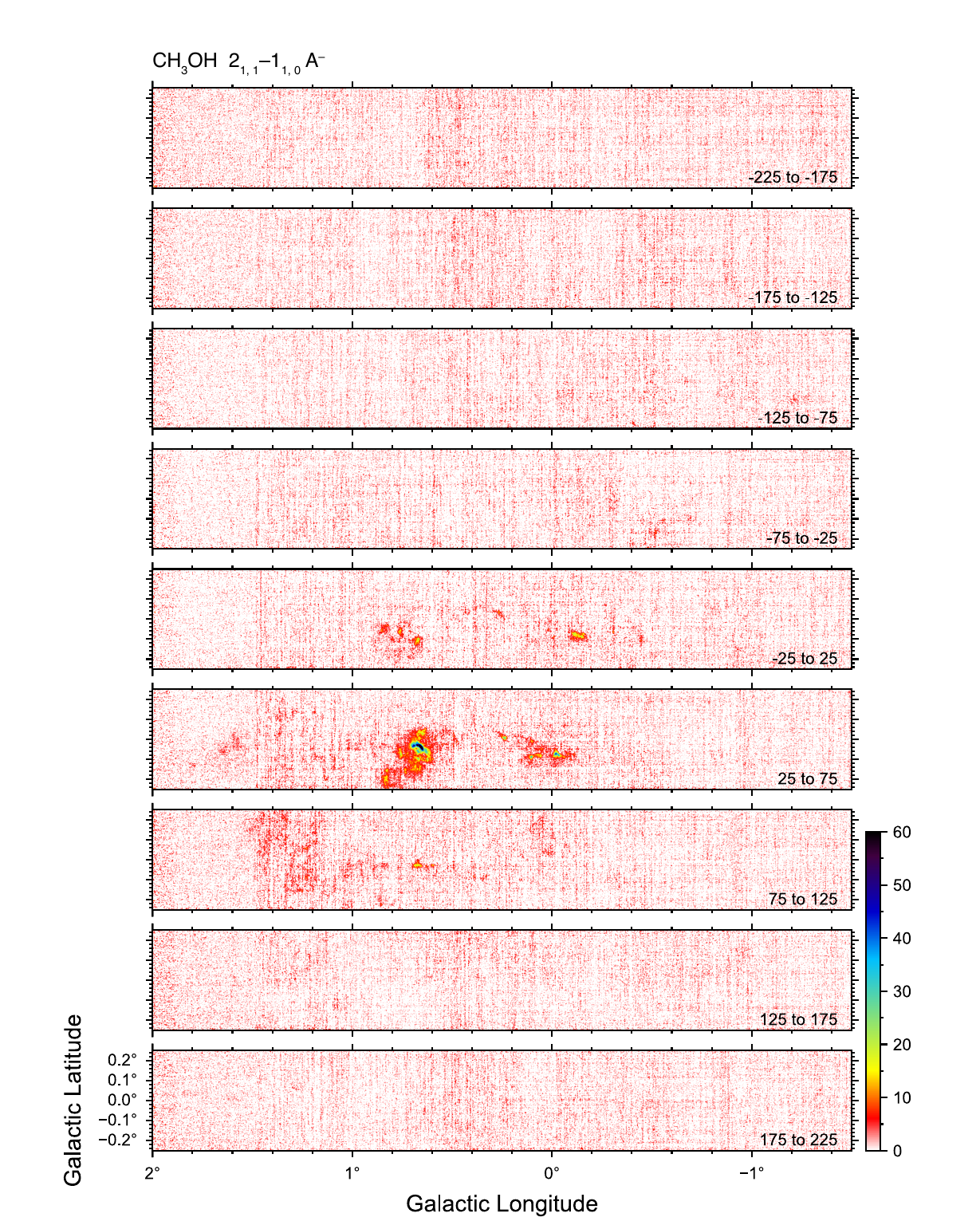}
\caption{Same as Figure~\ref{fig:h13cop_chmap}, but for the CH$_3$OH $2_{1,1}$--$1_{1,0}$ A$^-$ line.
\label{fig:ch3oh_chmap}}
\end{figure}

\begin{figure}[p!]
\figurenum{B5}
\centering
\includegraphics[height=0.78\textheight,keepaspectratio]{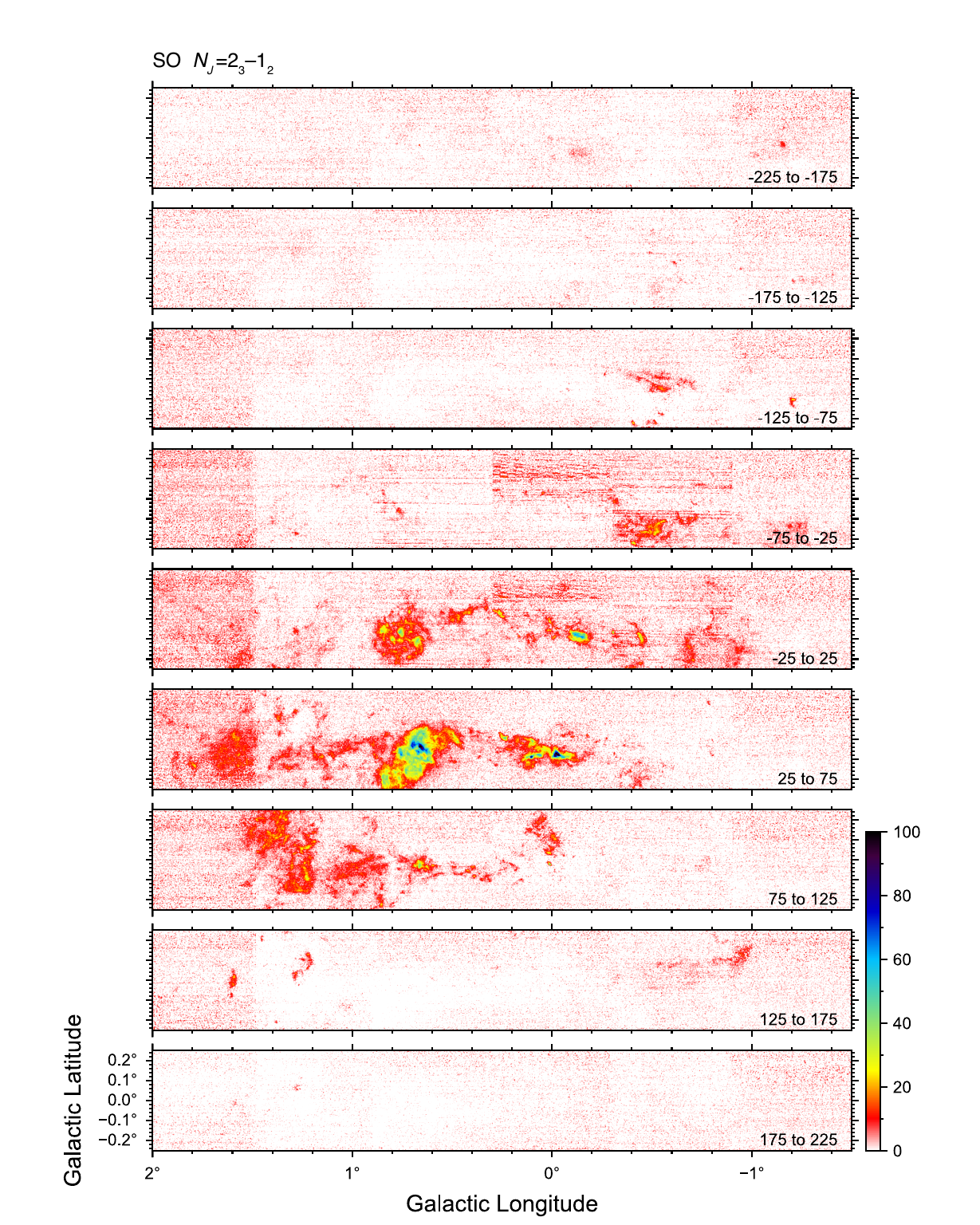}
\caption{Same as Figure~\ref{fig:h13cop_chmap}, but for the SO $N_J=2_3$--$1_2$ line.
\label{fig:so_chmap}}
\end{figure}

\begin{figure}[p!]
\figurenum{B6}
\centering
\includegraphics[height=0.78\textheight,keepaspectratio]{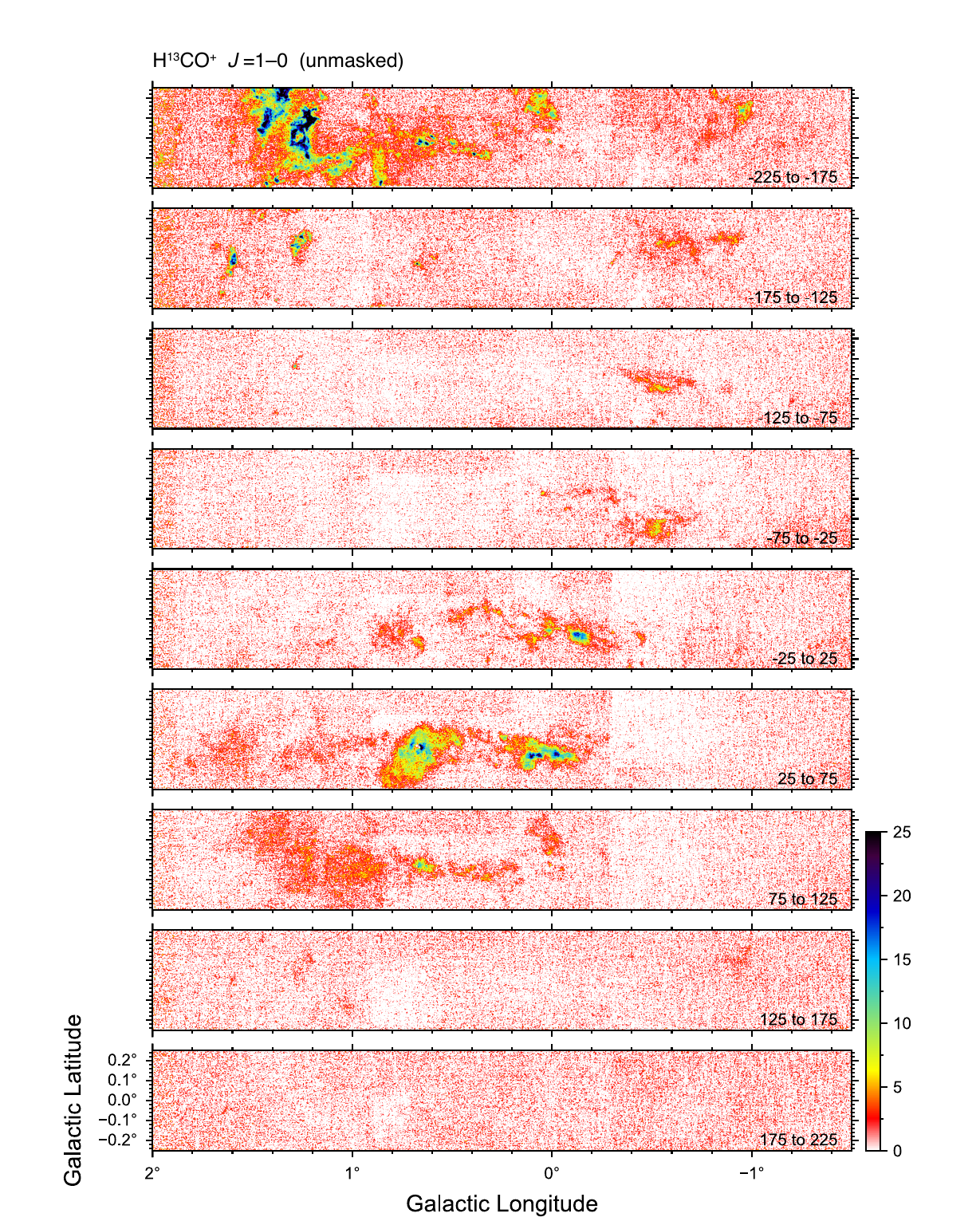}
\caption{Same as Figure~\ref{fig:h13cop_chmap}, but for the unmasked H$^{13}$CO$^+$ $J$=1--0 cube.
This figure is provided to show the unmasked cube before applying the CS-based mask.
\label{fig:h13cop_unmasked_chmap}}
\end{figure}

\begin{figure}[p!]
\figurenum{B7}
\centering
\includegraphics[height=0.78\textheight,keepaspectratio]{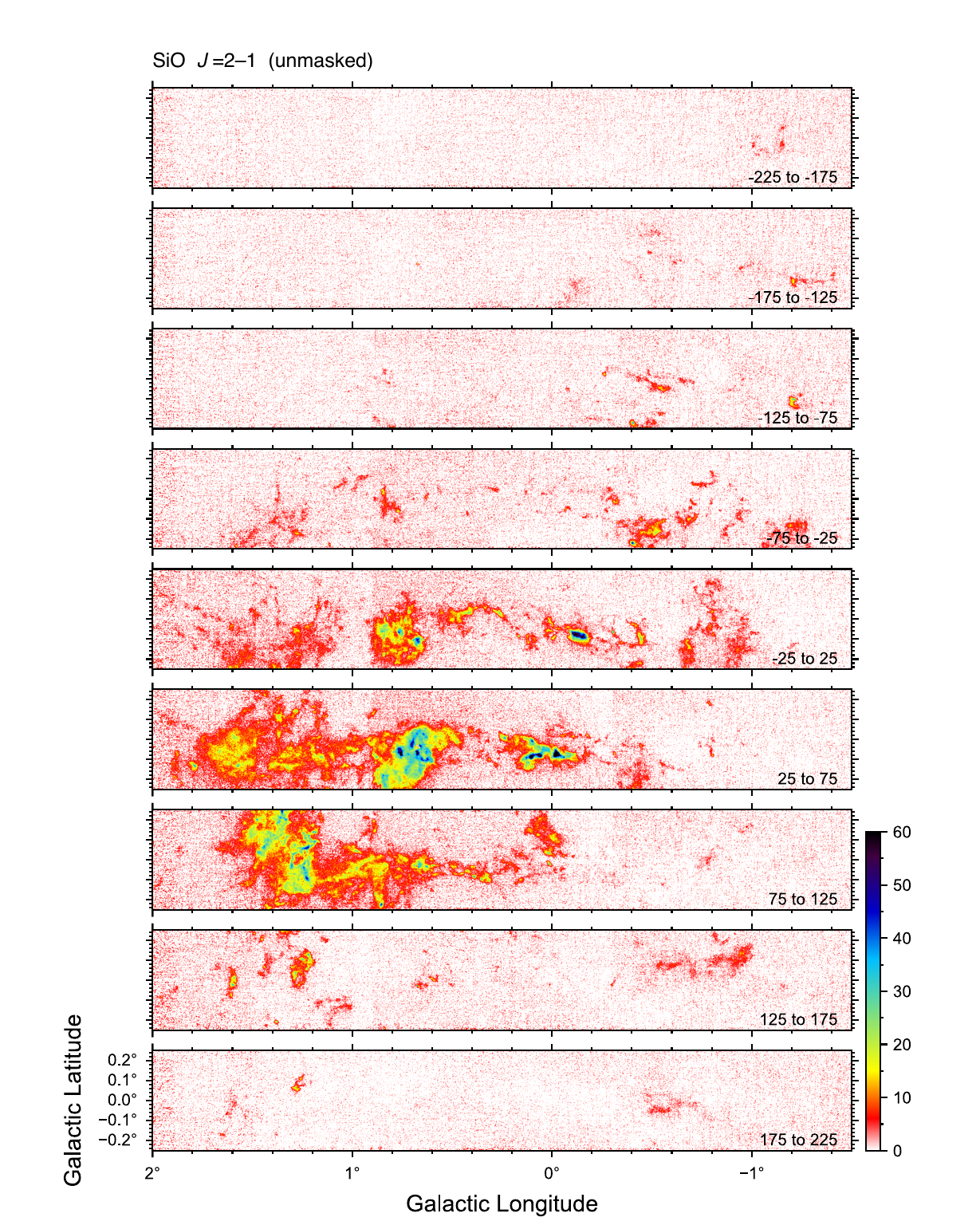}
\caption{Same as Figure~\ref{fig:h13cop_chmap}, but for the unmasked SiO $J$=2--1 cube.
This figure is provided to show the unmasked cube before applying the CS-based mask.
\label{fig:sio_unmasked_chmap}}
\end{figure}

Figures~\ref{fig:h13cop_chmap}--\ref{fig:so_chmap} show velocity-channel maps of the observed lines not shown in the main text.
They are constructed from the calibrated FITS cubes using the same 50~\kms{} intervals as the channel maps in Section~\ref{sec:chanmaps}, covering $-225$ to $+225$~\kms{}.
Figures~\ref{fig:h13cop_unmasked_chmap} and \ref{fig:sio_unmasked_chmap} show the unmasked H$^{13}$CO$^+$ and SiO channel maps.
Because the rest frequencies of H$^{13}$CO$^+$ and SiO separate by only about 320~\kms{}, emission from one line can appear in the velocity coverage of the other when the cubes are displayed without the CS-based mask.

\begin{figure}[p!]
\figurenum{B8}
\centering
\includegraphics[width=0.96\textwidth,keepaspectratio]{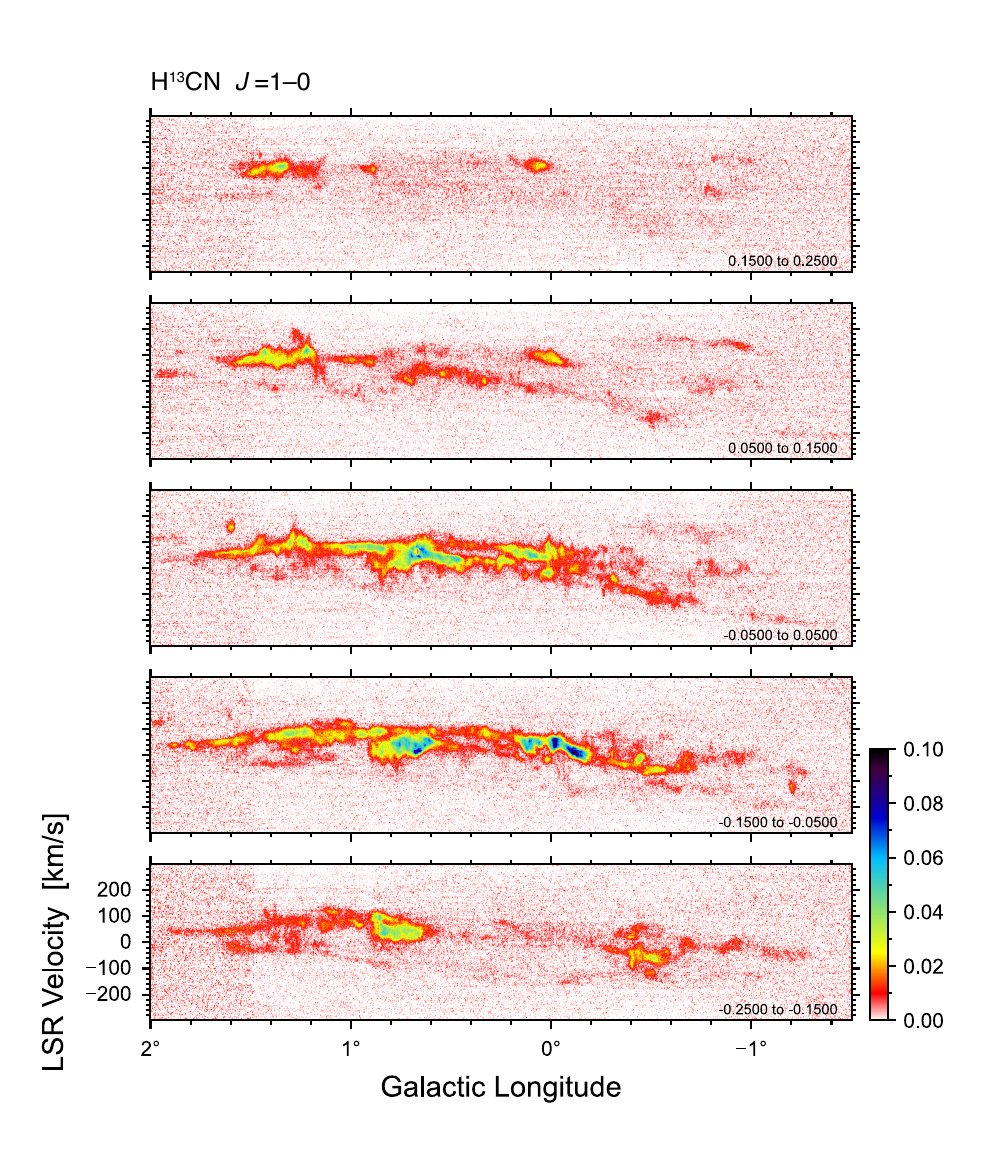}
\caption{Longitude--velocity maps of the H$^{13}$CN $J$=1--0 line.
Each panel shows emission averaged over a 0\fdg1-wide latitude bin.
The five panels span $-0\fdg25 \leq b \leq +0\fdg25$.
\label{fig:h13cn_lvmap}}
\end{figure}

\begin{figure}[p!]
\figurenum{B9}
\centering
\includegraphics[width=0.96\textwidth,keepaspectratio]{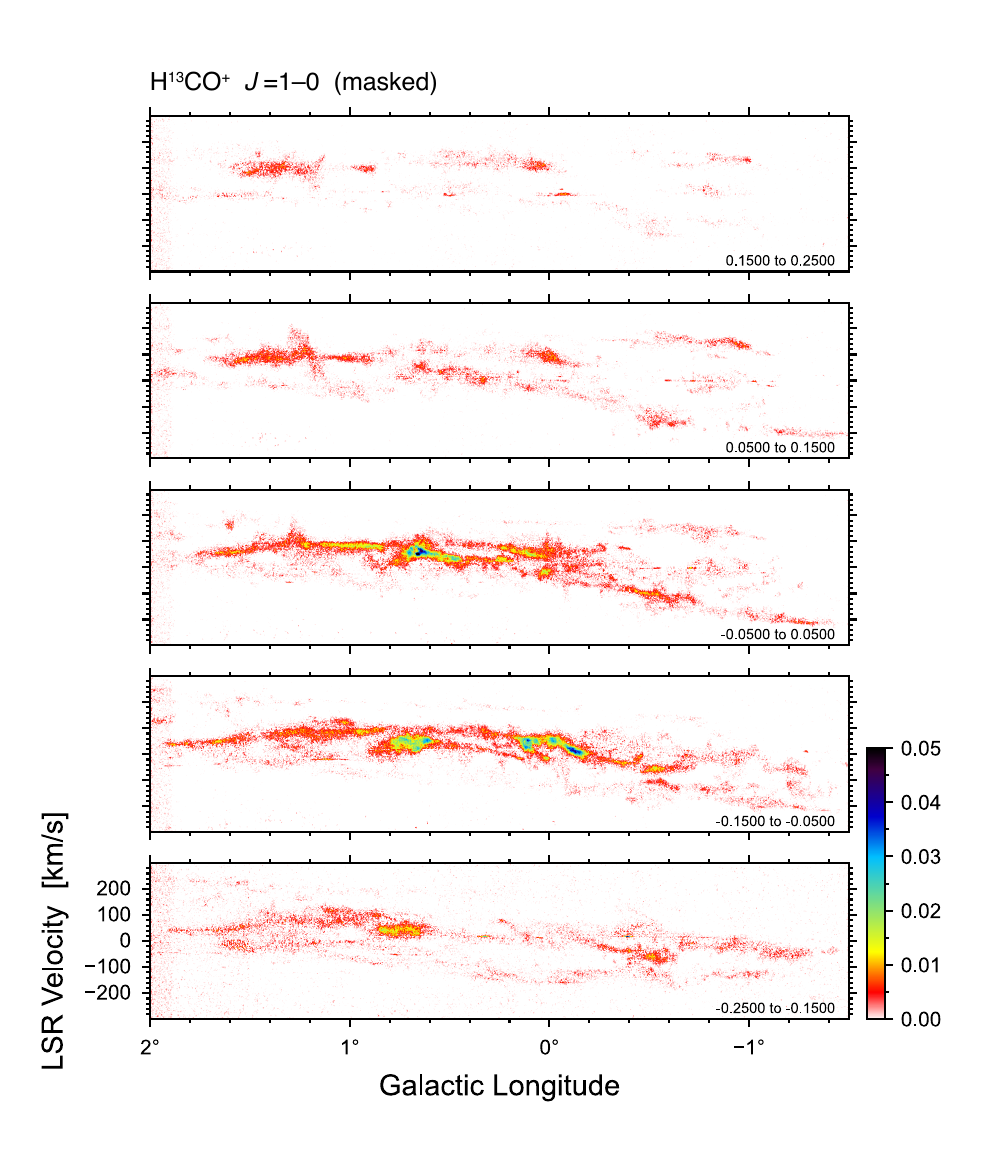}
\caption{Same as Figure~\ref{fig:h13cn_lvmap}, but for the masked H$^{13}$CO$^+$ $J$=1--0 cube.
\label{fig:h13cop_lvmap}}
\end{figure}

\begin{figure}[p!]
\figurenum{B10}
\centering
\includegraphics[width=0.96\textwidth,keepaspectratio]{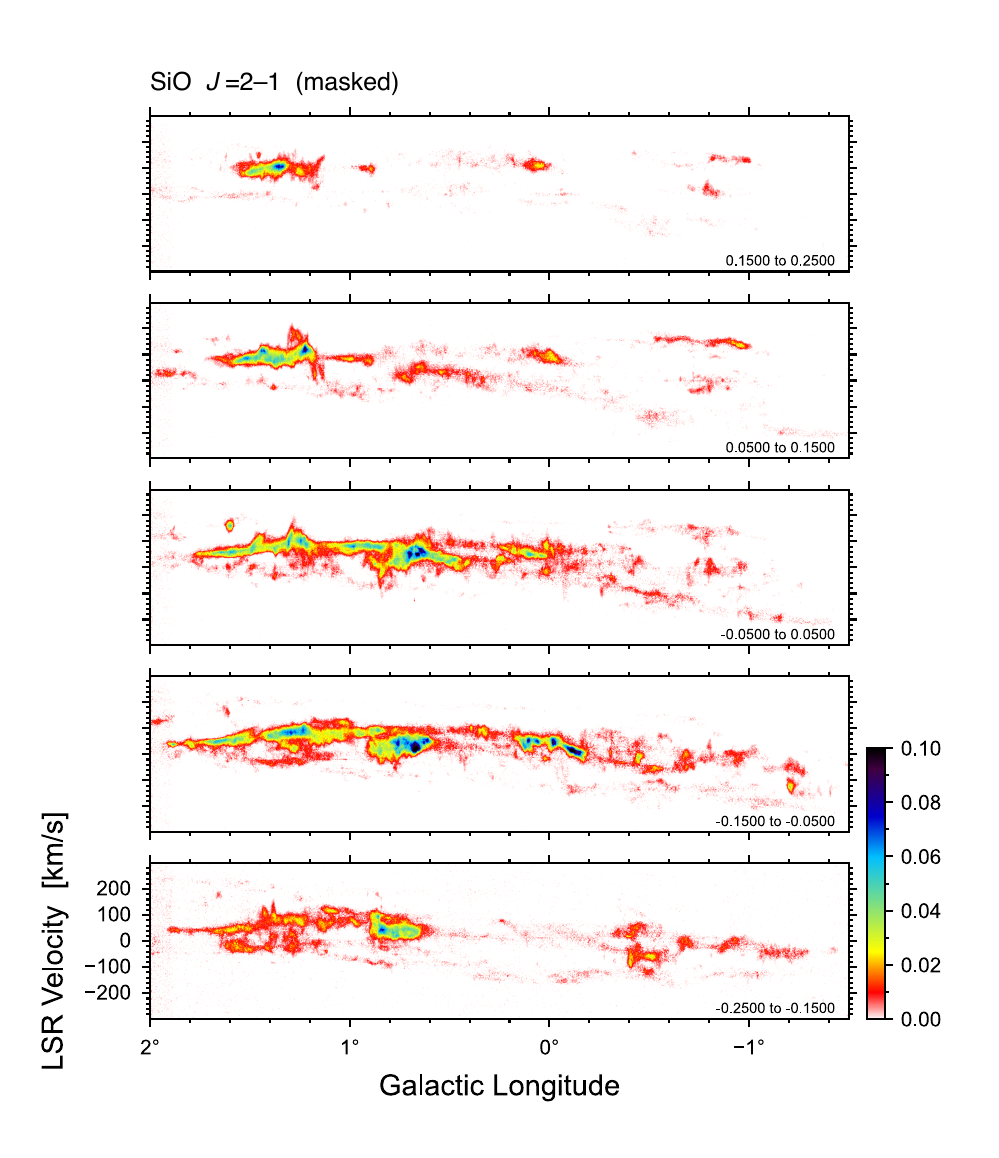}
\caption{Same as Figure~\ref{fig:h13cn_lvmap}, but for the masked SiO $J$=2--1 cube.
\label{fig:sio_lvmap}}
\end{figure}

\begin{figure}[p!]
\figurenum{B11}
\centering
\includegraphics[width=0.96\textwidth,keepaspectratio]{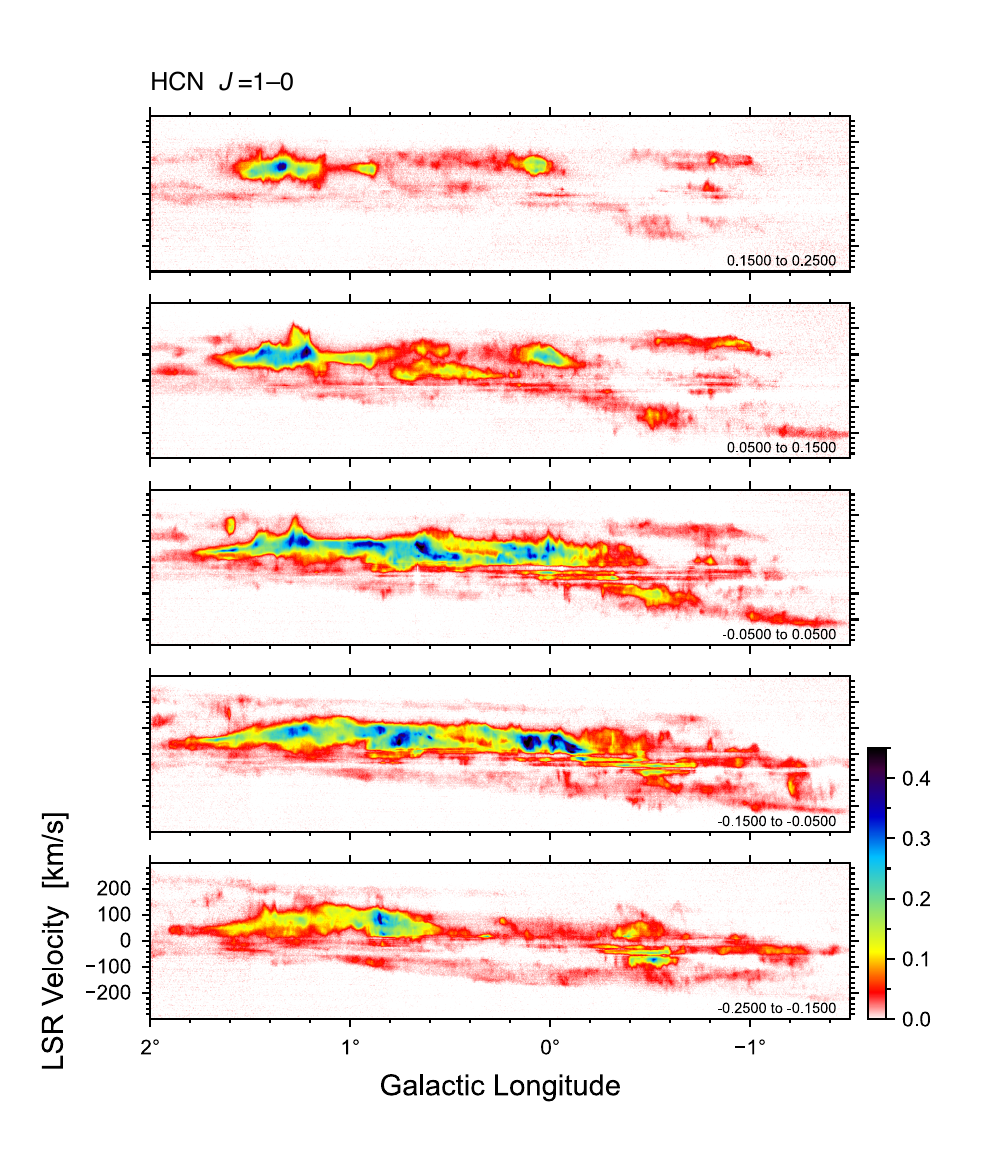}
\caption{Same as Figure~\ref{fig:h13cn_lvmap}, but for the HCN $J$=1--0 line.
\label{fig:hcn_lvmap}}
\end{figure}

\begin{figure}[p!]
\figurenum{B12}
\centering
\includegraphics[width=0.96\textwidth,keepaspectratio]{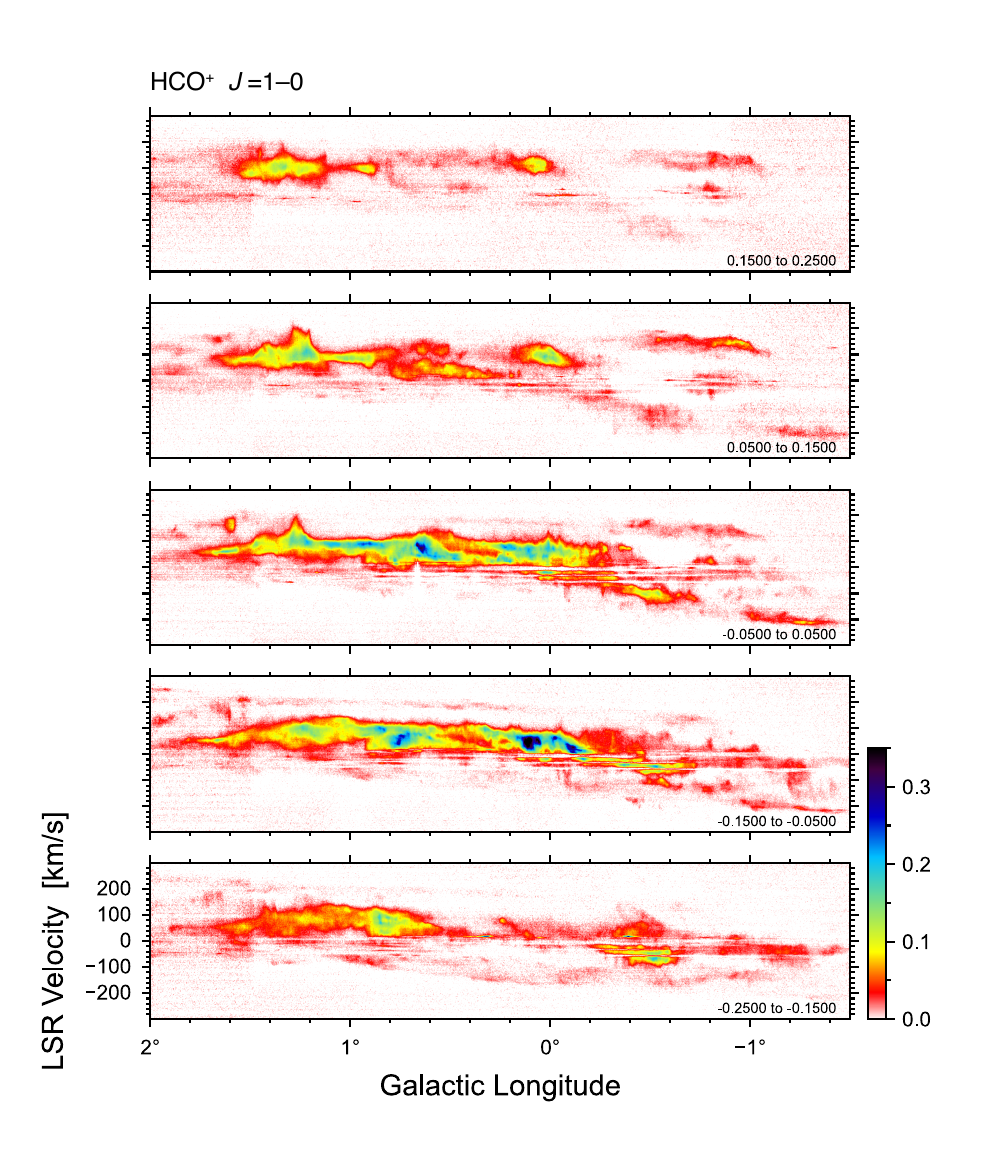}
\caption{Same as Figure~\ref{fig:h13cn_lvmap}, but for the HCO$^+$ $J$=1--0 line.
\label{fig:hcop_lvmap}}
\end{figure}

\begin{figure}[p!]
\figurenum{B13}
\centering
\includegraphics[width=0.96\textwidth,keepaspectratio]{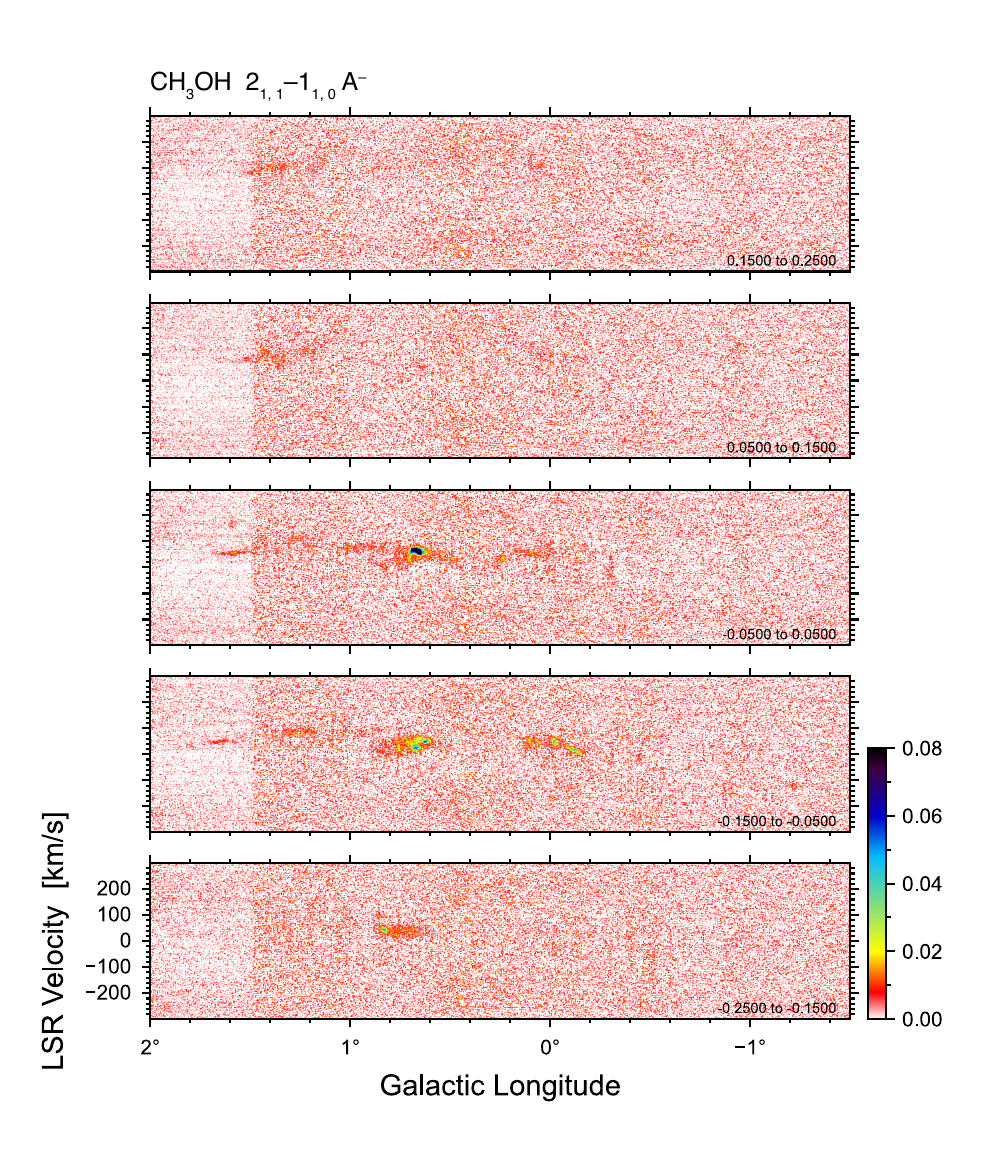}
\caption{Same as Figure~\ref{fig:h13cn_lvmap}, but for the CH$_3$OH $2_{1,1}$--$1_{1,0}$ A$^-$ line.
\label{fig:ch3oh_lvmap}}
\end{figure}

\begin{figure}[p!]
\figurenum{B14}
\centering
\includegraphics[width=0.96\textwidth,keepaspectratio]{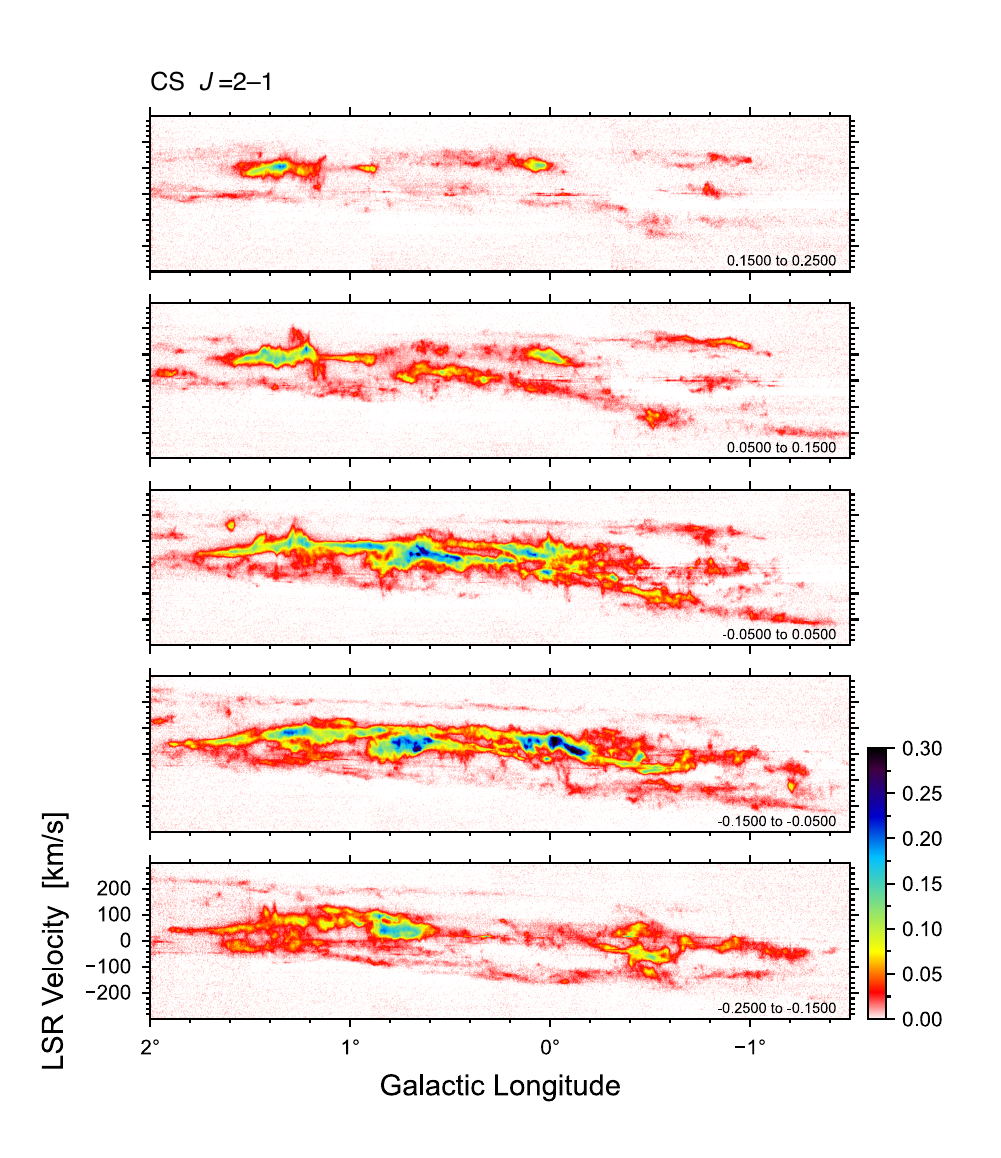}
\caption{Same as Figure~\ref{fig:h13cn_lvmap}, but for the CS $J$=2--1 line.
\label{fig:cs_lvmap}}
\end{figure}

\begin{figure}[p!]
\figurenum{B15}
\centering
\includegraphics[width=0.96\textwidth,keepaspectratio]{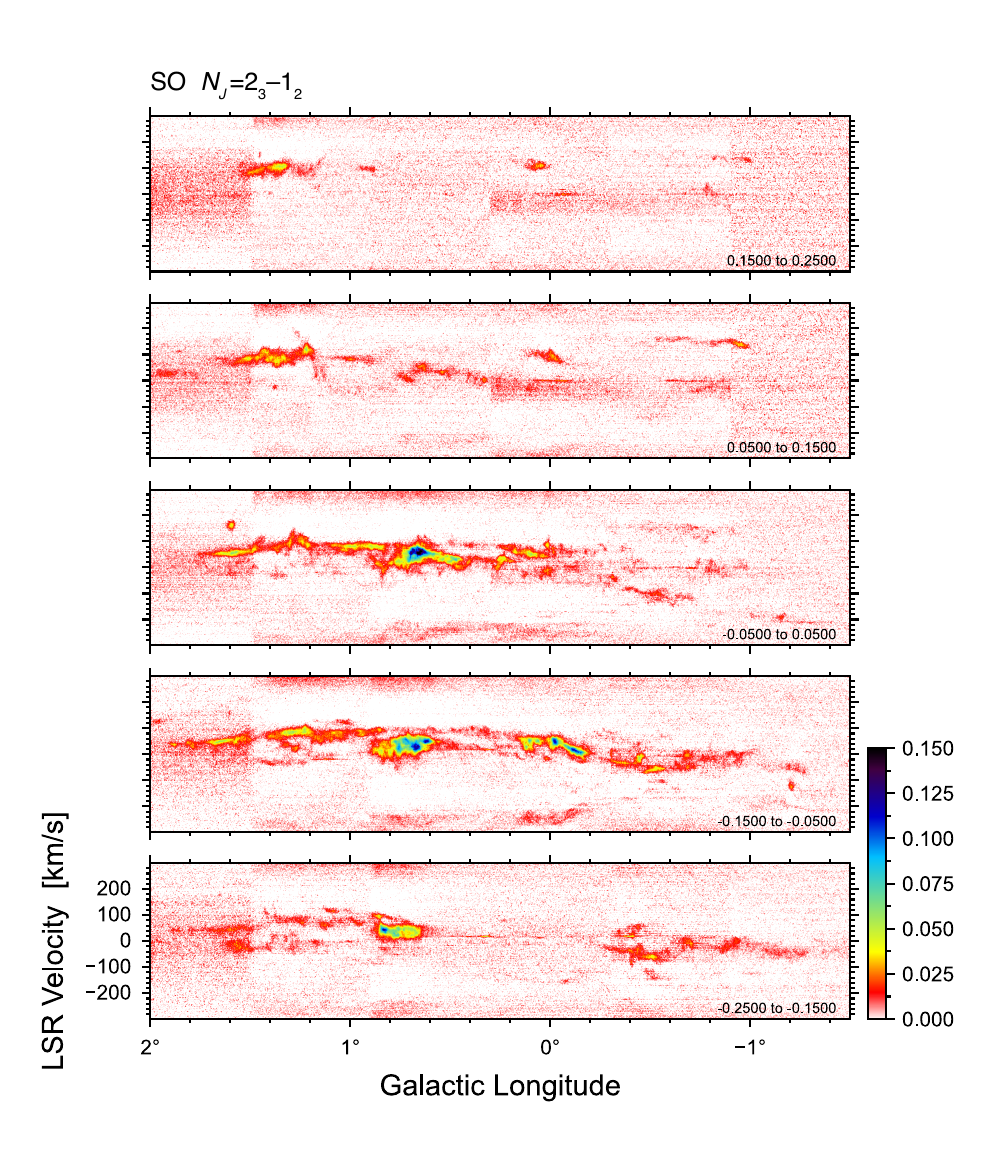}
\caption{Same as Figure~\ref{fig:h13cn_lvmap}, but for the SO $N_J=2_3$--$1_2$ line.
\label{fig:so_lvmap}}
\end{figure}

Figures~\ref{fig:h13cn_lvmap}--\ref{fig:so_lvmap} show longitude--velocity maps averaged over 0\fdg1 latitudes for all eight observed transitions, spanning $-0\fdg25 \leq b \leq +0\fdg25$.
These maps provide a complementary view of the velocity structure at different Galactic latitudes.
The SiO and H$^{13}$CO$^+$ longitude--velocity maps are constructed from the masked cubes.

\clearpage

\clearpage

\clearpage

\clearpage

\clearpage

\clearpage

\clearpage

\clearpage

\clearpage

\clearpage

\clearpage

\clearpage

\clearpage

\clearpage

\clearpage
\section{Integrated Line Intensities for the SiO-emitting Clouds} \label{sec:catalogproducts}

Table~\ref{tab:intensities} lists the integrated intensities of the eight observed lines for the SiO-emitting clouds, measured within the SCIMES masks.
The full table is available in a machine-readable format online.

\begin{deluxetable}{lccccccccccc}[ht!]
\tabletypesize{\scriptsize}
\tablenum{C1}
\tablecaption{Integrated Intensities of the Eight Observed Lines for the SiO-emitting Clouds \label{tab:intensities}}
\tablewidth{0pt}
\tablehead{
\colhead{ID} & \colhead{$l$} & \colhead{$b$} & \colhead{$V_{\rm LSR}$} & \colhead{$W_{\rm SiO}$} & \colhead{$W_{\rm H^{13}CN}$} & \colhead{$W_{\rm CS}$} & \colhead{$W_{\rm H^{13}CO^+}$} & \colhead{$W_{\rm HCN}$} & \colhead{$W_{\rm HCO^+}$} & \colhead{$W_{\rm SO}$} & \colhead{$W_{\rm CH_3OH}$} \\
\colhead{} & \colhead{(deg)} & \colhead{(deg)} & \colhead{(\kms{})} & \multicolumn{8}{c}{(K~\kms{})}
}
\startdata
1 & $-1.20$ & $-0.11$ & $-120$ & $4281$ & $2186$ & $10108$ & $262$ & $12631$ & $2472$ & $2569$ & $262$ \\
2 & $-1.19$ & $-0.13$ & $-41$ & $406$ & $146$ & $836$ & $31$ & $582$ & $405$ & $307$ & $5$ \\
3 & $-1.19$ & $-0.13$ & $-93$ & $97$ & $33$ & $178$ & $4$ & $118$ & $37$ & $33$ & $2$ \\
4 & $-0.96$ & $+0.14$ & $132$ & $2332$ & $807$ & $4431$ & $319$ & $3318$ & $2910$ & $1998$ & $160$ \\
5 & $-0.94$ & $-0.05$ & $-178$ & $39$ & $27$ & $104$ & $3$ & $37$ & $15$ & $22$ & $4$ \\
$\vdots$ & & & & & & & & & & & \\
\enddata
\tablecomments{Integrated intensities are obtained by summing $T_{\rm MB}\,\Delta v$ over all voxels within the SCIMES mask of each cloud, where $\Delta v = 2$~\kms{} is the velocity-channel width of the data cubes. Negative values are retained when the measured cloud-integrated intensity is nonpositive because of noise or baseline residuals. (This table is available in its entirety in machine-readable form in the online article.)}
\end{deluxetable}

\section{\texorpdfstring{H$^{13}$CN Optical Depth, Excitation Temperature, and Column Density}{H13CN Optical Depth, Excitation Temperature, and Column Density}} \label{sec:lteform}

This appendix summarizes the equations used to estimate the effective H$^{13}$CN optical depth, the effective excitation temperature, and the LTE H$^{13}$CN column density.
All quantities are evaluated for the cloud-integrated intensities measured within the SCIMES masks.
For a cloud $c$, the integrated intensity of line $i$ is
\begin{equation}
 W_{i,c} =
\sum_{(l,b,v)\in c} T_{{\rm MB},i}(l,b,v)\,\Delta v ,
\end{equation}
where $i$ denotes HCN or H$^{13}$CN and $\Delta v=2$~\kms{} is the channel width.
Assuming a common excitation temperature for the two isotopologues, the cloud-integrated intensity ratio is
\begin{equation}
 R_c =
\frac{W({\rm HCN})}{W({\rm H^{13}CN})}
=
\frac{1-\exp\left(-X_{12/13}\tau_{13,c}\right)}{1-\exp\left(-\tau_{13,c}\right)},
\end{equation}
where $X_{12/13}=24$ is the adopted $^{12}$C/$^{13}$C ratio and $\tau_{13,c}$ is the effective optical depth of H$^{13}$CN.
We solve this equation numerically for clouds with $1<R_c<X_{12/13}$.

We define the cloud-averaged H$^{13}$CN brightness temperature as
\begin{equation}
 \langle T_{{\rm MB},13}\rangle_c =
\frac{W_{13,c}}{N_{{\rm vox},c}\,\Delta v},
\end{equation}
where $N_{{\rm vox},c}$ is the number of voxels in the SCIMES mask.
Assuming uniform excitation and optical depth within each cloud, the standard radiative-transfer relation gives
\begin{equation}
 \langle T_{{\rm MB},13}\rangle_c =
\left[J_\nu(T_{\rm ex}) - J_\nu(T_{\rm bg})\right]
\left[1-\exp(-\tau_{13,c})\right].
\end{equation}
Here the beam-filling factor is set to unity, so the resulting $T_{\rm ex}$ should be interpreted as a beam- and mask-averaged effective excitation temperature rather than an intrinsic excitation temperature.
The radiation temperature function is
\begin{equation}
 J_\nu(T)=
\frac{h\nu/k_{\rm B}}{\exp(h\nu/k_{\rm B}T)-1},
\end{equation}
where $h$ is the Planck constant, $k_{\rm B}$ is the Boltzmann constant, and $\nu$ is the line frequency.
Thus,
\begin{equation}
 J_\nu(T_{\rm ex}) =
\frac{\langle T_{{\rm MB},13}\rangle_c}{1-\exp(-\tau_{13,c})}
+ J_\nu(T_{\rm bg}),
\end{equation}
which is inverted to obtain the effective $T_{\rm ex}$ for each cloud.
We use this effective value only to justify adopting $T_{\rm ex}=5$~K when calculating the H$^{13}$CN column density.

For a transition from upper level $u$ to lower level $l$, the total column density under LTE is written as
\begin{equation}
 \begin{array}{l}
N_{\rm tot} =
\frac{8 \pi k_{\rm B} \nu^2}{h c^3 A_{ul}}
\frac{Q(T_{\rm ex})}{g_u}
\exp\!\left(\frac{E_u}{k_{\rm B}T_{\rm ex}}\right) \\
\qquad\qquad \times
\frac{J_\nu(T_{\rm ex})}{J_\nu(T_{\rm ex})-J_\nu(T_{\rm bg})}
C_\tau
\int T_{\rm MB}\,dv ,
\end{array}
\end{equation}
where $A_{ul}$ is the Einstein coefficient, $Q(T_{\rm ex})$ is the partition function, $g_u$ is the upper-state degeneracy, $E_u$ is the upper-state energy, and $T_{\rm bg}=2.725$~K.
The optical-depth correction factor is
\begin{equation}
 C_\tau = \frac{\tau}{1-\exp(-\tau)} .
\end{equation}
For the cloud mass estimates in the main text, we use the optically thin limit for H$^{13}$CN, $C_\tau=1$, with $T_{\rm ex}=5$~K.

\section{SiO Ratios versus Cloud-Averaged Density} \label{sec:ratio_n}

\begin{figure}[ht!]
\figurenum{E1}
\plotone{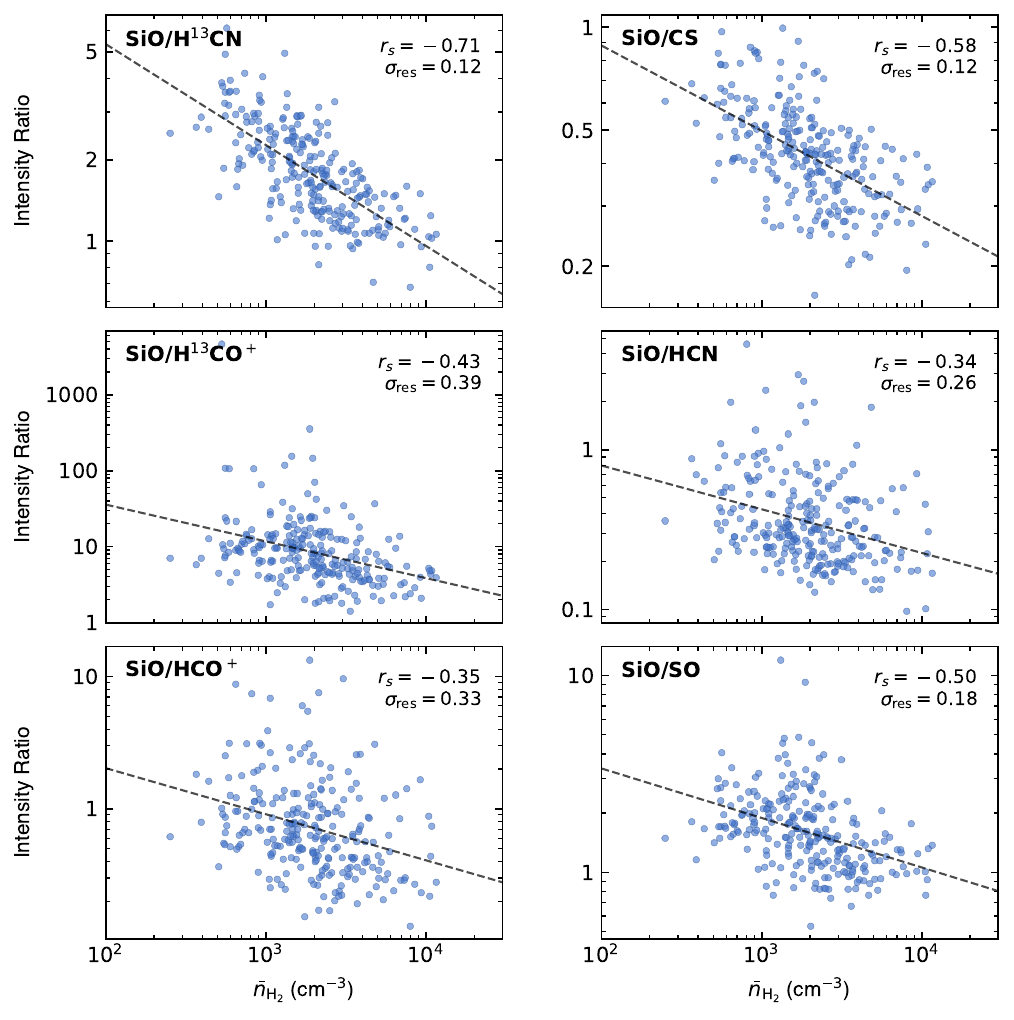}
\caption{Same as Figure~\ref{fig:main}, but with $\nbar{}$ on the horizontal axis instead of $\ntdyn{}$. \label{fig:ratio_n}}
\end{figure}

Figure~\ref{fig:ratio_n} shows the same six SiO intensity ratios as in Figure~\ref{fig:main}, but plotted against $\nbar{}$ instead of $\ntdyn{}$.
Compared with Figure~\ref{fig:main}, the correlations are systematically broader and less tight, as quantified in Table~\ref{tab:spearman}.

\clearpage

 %% ============================================================
%% References
%% ============================================================

\twocolumngrid

\end{document}